\documentclass[twocolumn]{autart}  

\usepackage{graphicx}     

\usepackage[active]{srcltx}

\usepackage{verbatim}
\usepackage{epsfig}
\usepackage{amsmath}
\usepackage{amssymb}
\usepackage{psfrag}
\usepackage[T1]{fontenc}
\usepackage{graphicx}
\usepackage[usenames, dvipsnames]{pstricks}
\usepackage{pst-grad} 
\usepackage{pst-plot} 
\usepackage{algorithm}
\usepackage{algpseudocode}
\usepackage{url}
\usepackage{setspace}
\usepackage{mathtools}
\usepackage{subcaption}

\graphicspath{{./figure/}}

\newtheorem{definition}{Definition}

\newtheorem{remark}{Remark}

\newcommand{\squeezeup}{\vspace{-2.5mm}}

\DeclareMathOperator*{\argmax}{argmax}

\begin{document}

\begin{frontmatter}

\title{Physics Informed Topology Learning in Networks of Linear Dynamical Systems} 


\author[Paestum]{Saurav Talukdar}\ead{sauravtalukdar@umn.edu},  
\author[Rome]{Deepjyoti Deka}\ead{deepjyoti@lanl.gov},
\author[Paestum]{Harish Doddi}\ead{doddi003@umn.edu},
\author[Baiae]{Donatello Materassi}\ead{dmateras@utk.edu},  
\author[Rome]{Misha Chertkov}\ead{chertkov@lanl.gov},
\author[Paestum]{Murti V. Salapaka}\ead{murtis@umn.edu}

\address[Paestum]{University of Minnesota Twin Cities, Minneapolis, USA} \address[Rome]{Los Alamos National Laboratory, Los Alamos, USA}\address[Baiae]{University of Tennessee, Knoxville, USA}    


\begin{abstract}             
Learning influence pathways of a network of dynamically related processes from observations is of considerable importance in many disciplines. In this article, influence networks of agents which interact dynamically via linear dependencies are considered. An algorithm for the reconstruction of the topology of interaction based on multivariate Wiener filtering is analyzed. It is shown that for a vast and important class of interactions, that respect flow conservation, the topology of the interactions can be {\it exactly} recovered. The class of problems where reconstruction is guaranteed to be exact includes power distribution networks, dynamic thermal networks and consensus networks. The efficacy of the approach is illustrated through simulation and experiments on consensus networks, IEEE power distribution networks and thermal dynamics of buildings.
\end{abstract}

\end{frontmatter}

\section{Introduction}
Networks are widely used to represent the functioning of complex physical systems like the power grid \cite{deka2017structure}, brain \cite{bullmore2009complex} as well nonphysical systems like social relationships \cite{scott2017social}, financial systems \cite{allen2009networks} and many others. In some applications, there is a clear and evident physical network of agents. For example, consider a network where agents comprise wireless hubs that receive and transmit signals according to a network topology. In other applications, a physical network is not evident, for example, sensors measuring temperature at various zones in a building, do not admit easy identification of physical links that connect measurements. In either scenario, network models play important role for determining influences and identification of important clusters. The resulting influence pathways can suggest methods to steer the system toward a desired behavior. We note that for applications where an actual physical network of interactions is evident, the network of mutual influences can differ from its physical network, and often provides complimentary information. In many applications it is possible to manipulate the system to help identifying the network. An active approach for identifying the presence of influences between agents may entail removing agents from the network to evaluate their impact on other agents and then infer influence pathways \cite{nabi2014network}. Such approaches are invasive. In many applications it may not be possible to excite/perturb the system to decipher the influence of one variable on the rest. For instance, in the stock market, it is not possible to set the price of a particular stock to evaluate the impact on the other stocks. Thus, there is a clear need of non invasive approaches to infer the network representation of complex systems \cite{materassi2015identification}.

Non-invasive methods of learning the network topology is an active area of research in multiple disciplines including control theory, computer science and statistics to name a few.
Here an initial focus utilized modeling activity of agents via random variables; the graphical models approach (see \cite{Pea88},\cite{pearl2009causality}, \cite{meinshausen2006high}, \cite{friedman2008sparse}) employed the random variable based abstraction extensively. More recently, network topology learning methods have utilized time series measurements of agents' activity modeled as stochastic processes \cite{dahlhaus2000graphical}, \cite{gonccalves2008necessary}.
The approach of using stochastic processes to model time-series is well-suited for scenarios where agents interact dynamically; where the past or present state of an agent can affect the present state of another agent. Such interdependence is particularly apt for modeling high resolution time-series measurements, which is becoming more widespread due the increasing availability of smart sensors with high bandwidth (for example, Phase Measurement Units employed by the power grid \cite{nuqui2005phasor}) in many domains including Internet of Things.


In \cite{materassi2012problem} the authors present a multivariate Wiener filtering for inferring the network structure of agents that interact via linear dynamical relations. If interactions between the present state of an agent with the present state of another agent exists in the network, then inferred topology is shown to indicate only dependencies and does not provide causal characterization of the inferred edges. However, in the case where the present state of an agent depends only on the strict past of other agents of the network, it is shown that the network structure can be inferred exactly along with causal characterization of the inferred edges. For nonlinear systems with a known bound on the in degree of a node, the authors in \cite{quinn2015directed} use a directed information approach to infer directed graphs. The assumption of strictly causal dynamics is indeed a significant enabling factor to guarantee exact topology inference with edge directions in \cite{materassi2012problem} and \cite{quinn2015directed} .
In \cite{shahrampour2015topology} the authors use a power spectral analysis approach to infer the network topology. For consensus dynamics, a decentralized and distributed topology learning scheme is presented in \cite{kibangou2012decentralized} and \cite{morbidi2014distributed} respectively. The works summarized above, primarily use a mix of signal processing or optimization schemes or structural restrictions like radial topology \cite{talukdar2017exact} to infer the topology. A crucial characteristic is that, none of the above works, utilize any knowledge about the underlying physics of the system toward topology learning.

 \textsl{Our Contribution}: In this article, we focus on topology learning in dynamical physical systems like power networks \cite{chowdhury2009microgrids} and thermal dynamic networks \cite{reynders2014quality}, where influences between agents, when present, are bi-directional representing a form of coupling and not necessarily a cause effect relationship. In these type of situations, the assumption of strict causality is not valid. We present a new algorithm for exact topology inference, which utilizes both the magnitude and phase response of multivariate Wiener filters for determining the presence/absence of links between two nodes. The main result establishes that the confounding effects of indirect effects of an agent on another can be detected using the phase information in the multivariate Wiener filters that estimate a time-series from the rest. We provide provable guarantees for consistency of the inferred topology with the underlying topology, without resorting to structural restrictions like the topology being radial \cite{talukdar2017learning} or bounded in-degree or strictly causal dynamics and without relying on information of system parameters or exogenous inputs. The consistency result holds even in the presence of feedback loops unlike \cite{materassi2013reconstruction}, \cite{talukdar2017learning}. More importantly, the algorithm is applicable even when the nodal exogenous inputs/noise are colored unlike prior work where assumptions of noise being white are necessary \cite{pereira2010learning}.
 Of particular focus in our work are physical flow networks like power networks, thermal dynamic networks where we present interesting connections of the phase response in our algorithm with physical conservation laws.

 An important application of topology learning is in real time monitoring of a network, which places a requirement of algorithms that can can infer the topology with finite/ limited samples of measurements from each node. The algorithm proposed in this article when used with conventional Wiener filtering, is guaranteed to recover the exact topology when sufficient data samples per node are available; here, particularly in the low sample regime the errors in topology inference are high. We introduce group Lasso \cite{bach2008consistency} based regularizers in the Wiener filtering optimization problem and demonstrate that our algorithm when used with regularized version of Wiener filtering provides accurate topology estimation even in the low sample regime. We demonstrate the benefit of using regularizers in topology learning for real time applications.
 The effectiveness of the algorithms and theory presented is illustrated through simulations on power distribution networks,  thermal dynamics of buildings and directed consensus networks.

Preliminary work subsumed by this article instantiated to various application domains have appeared in \cite{talukdar2017learning} for power grid networks, \cite{harish2018topology} for thermal dynamics of buildings and \cite{ctalukdar2017learning} for consensus networks. This article is a detailed version with complete proofs with a presentation from a general linear dynamical system perspective and explores connections with physical laws.

In the next section we present the framework, define the problem of topology learning and provide some motivating examples. Wiener filtering approach for topology inference is summarized in Section 3 followed by the derivation of the exact topology learning algorithm and its connections with physical conservation laws in Section 4. In Section 5, we illustrate the performance of the learning algorithm on power grid simulations, Energy plus based building simulation and Raspberry Pi based experimental data from consensus dynamics. Conclusions are presented in Section 6.

\section{Preliminaries}
Suppose $x_i$, $i= 1,\ldots,n$ represent $n$ measured time-series. Further assume that the dynamics generating the time-series $x_i$ satisfy,

\begin{align}\label{eqn:lindyn}
  \sum_{m=1}^{l}a_{m,i}\frac{d^mx_i}{dt^m}=\sum_{j=1,j\neq i}^{n}b_{ij}(x_j(t) - x_i(t)) + p_i(t),
\end{align}
$ i \in\{1,2,..,n\}$, where, the exogenous forcing $p_i(t)$ is a zero mean wide sense stationary (WSS) process uncorrelated with $p_j(t)$ for $j \neq i$. Note that a linear transformation of $\{p_j(k)\}_{j=1}^n$ results in $\{x_j(k)\}_{j=1}^n$. We assume that the above dynamics is stable. Thus, $\{x_j(k)\}_{j=1}^n$ are a collection of WSS processes and the collection $(\{x_j(k)\}_{j=1}^n,\{p_j(k)\}_{j=1}^n)$ are jointly wide sense stationary processes (JWSS) \cite{gubner2006probability}. Here, $x_i(t)\in \mathbb{R}$ is a state of the system, which is assumed to be measured, and, $a_{m,i}\in \mathbb{R}$, $b_{ij} \in \mathbb{R}_{\geq 0}$.
The $z$ transform of the discretization of continuous time dynamics (\ref{eqn:lindyn}) is given by,
\begin{align}\label{eqn:zdyn}
  S_i(z)X_i(z) = \sum_{j=1,j\neq i}^{n}b_{ij}X_j(z) + P_i(z),
\end{align}
where, $S_i(z)=\sum_{m=1}^{l}a_{m,i}((\frac{2(1-z^{-1})}{\Delta t(1+z^{-1})})^{m}+\sum_{j=1,j\neq i} c_{ij}$, is the $z$ domain operator determined by $x_i$ and its derivatives along with their coefficients in (\ref{eqn:lindyn}) and the sampling time $\Delta t$ (using Bilinear transform). Here, $X_i(z)$ is the $z$ transform of $x_i(k)$. Rewriting (\ref{eqn:zdyn}) we have,
\begin{align}\label{eqn:ldg}
  X_i(z) = \sum_{j=1,j\neq i}^{n}H_{ij}(z)X_j(z) + E_i(z)
\end{align}
where, $H_{ij}(z) = \frac{b_{ij}}{S_i(z)}, E_i(z) = \frac{1}{S_i(z)}P_i(z)$. It can be shown that $e_i(k)$ are uncorrelated with $e_j(k)$ for $j \neq i$, where, $e_i(k)$, is the inverse $z$ transform of $E_i(z)$, and is a zero mean wide sense stationary sequence for all $i= 1,...,n$. Summarizing, the network dynamics is represented as,
\begin{align}\label{eqn:netdyn}
  X(z)&= H(z)X(z) + E(z),~\text{where~},\\
  X(z)&=[X_1(z) \ X_2(z)\ ...\ X_n(z)]^T,\nonumber \\
  E(z)&=[E_1(z) \ E_2(z)\ ...\ E_n(z)]^T, H(z)(i,j) = H_{ij}(z).\nonumber
\end{align}
Note that the diagonal entries of the matrix $H(z)$ are 0. For well posedness we assume that $I-H(z)$ is invertible almost everywhere. Next, we associate a graphical representation derived from the transfer function matrix $H(z)$.

\begin{figure}[tb]
	\centering
	\begin{tabular}{ccc}
		\includegraphics[width=0.29\columnwidth]{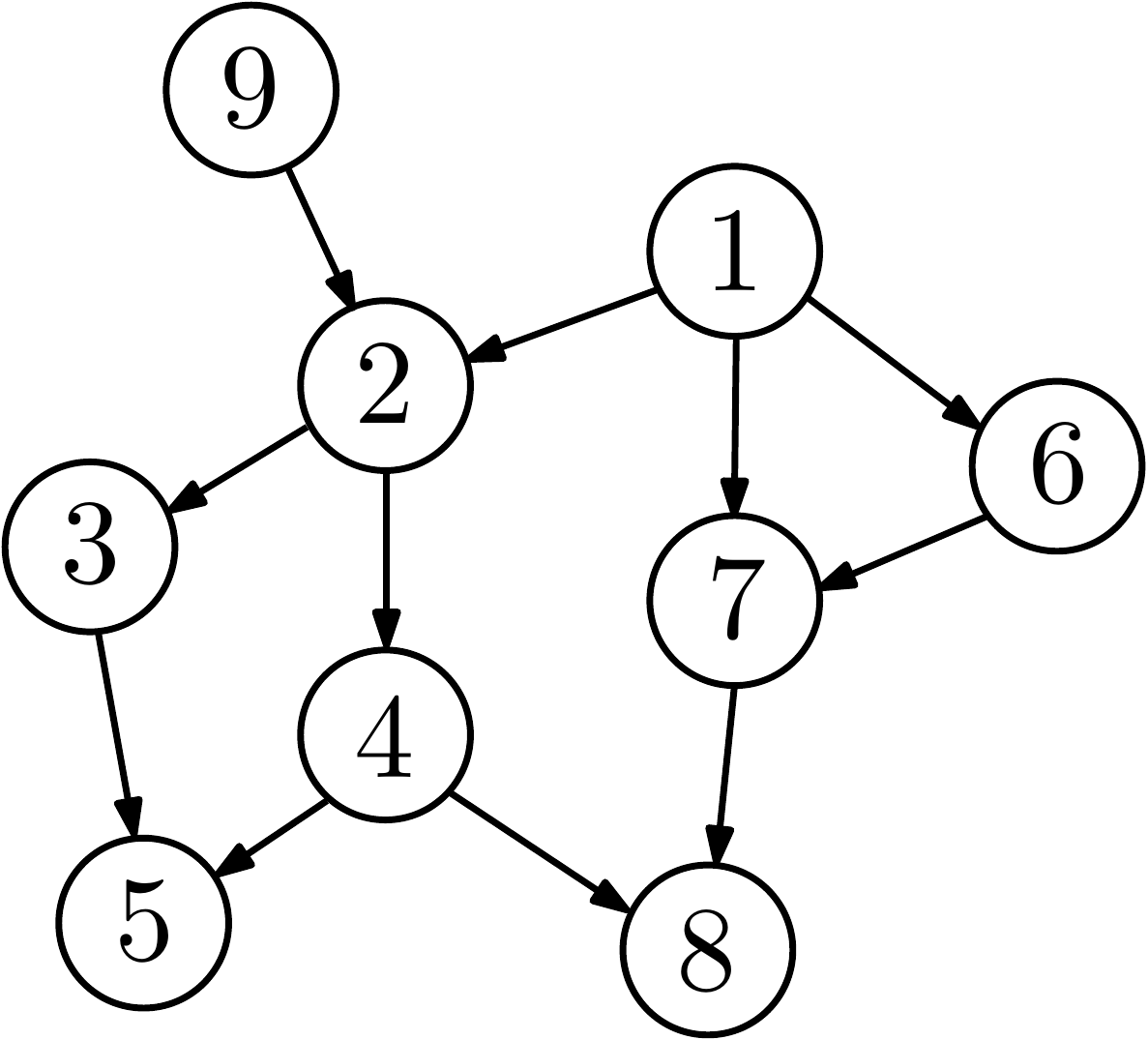} &
		\includegraphics[width=0.29\columnwidth]{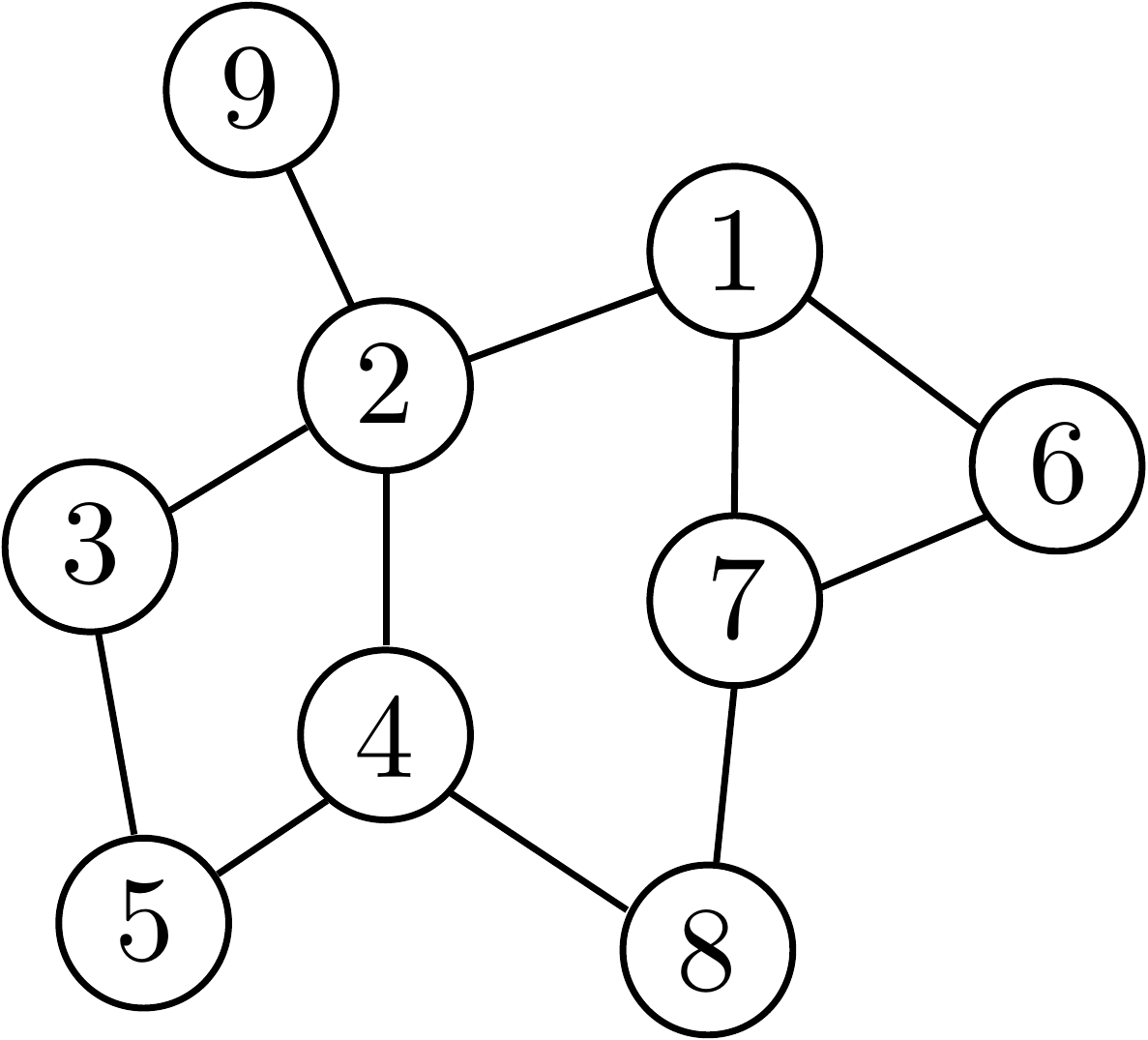} &
		\includegraphics[width=0.29\columnwidth]{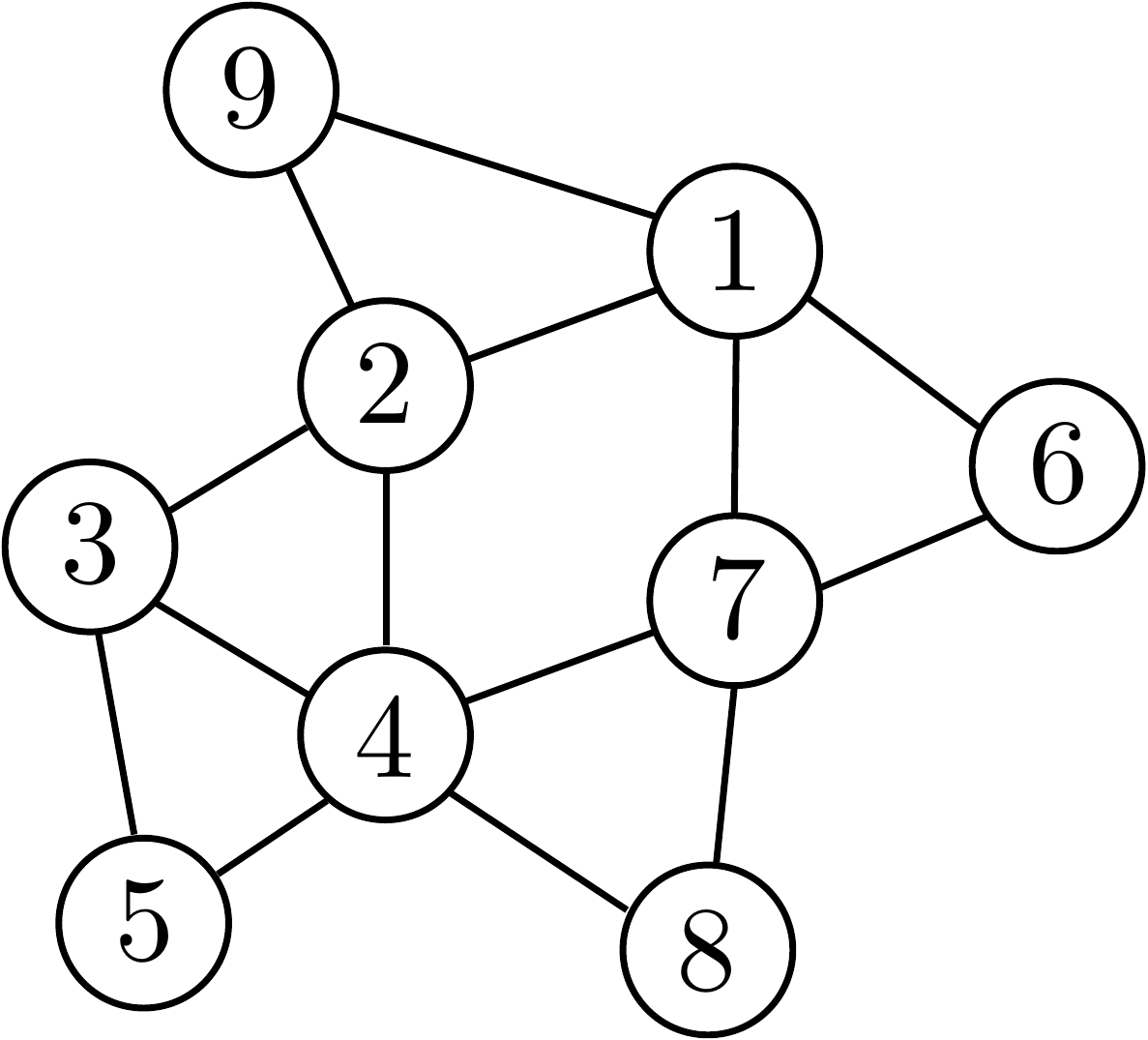}\\
		(a) & (b) & (c)
	\end{tabular}
	\caption{A generative graph $\mathcal{G}$ is shown in (a), its topology $\mathcal{G}_T$ in (b), and its moral graph, $\mathcal{G}_M$ in (c) \cite{materassi2012problem}.}
		\label{fig:topexample} 
\end{figure}

\textbf{Graphical Representation:} Consider a directed graph $\mathcal{G}=(\mathcal{V},\mathcal{E})$ with $\mathcal{V}=\{1,...,n\}$ being the nodes and $\mathcal{E} = \{(i,j)|H_{ij}(z) \neq 0\}$ as the edge set. Each node $i\in \mathcal{V}$ is a representation of the measured time series $x_i(k)$. We refer to the directed graph $\mathcal{G}$ to be the \emph{generative graph} of the measured time series. In the graph $\mathcal{G}$, $(i,j)$ denotes a directed edge from $j$ to $i$ if $H_{ij}(z) \neq 0$, where $j$ is referred as the parent of $i$ and $i$ is referred as the child of $j$. We will refer to nodes having common children as spouses of each other, for example $i \in \mathcal{V}$ and $j\in \mathcal{V}$ are spouses if there exist a node $k \in \mathcal{V}$ such that $H_{ki}(z) \neq 0$ and $H_{kj}(z) \neq 0$ almost surely. Let $C_j, P_j, K_j$ denote the set of nodes consisting of children, parents and spouses of node $j$ in the generative graph $\mathcal{G}$. For illustration, in Fig. \ref{fig:topexample} we show a generative graph, where, nodes $1$ and $9$ are parents of node $2$, while, node $2$ is the child of nodes $1$ and $9$. Here, nodes $1$ and $9$ are spouses of each other. Given a generative graph $\mathcal{G}$, its \emph{topology} is defined as the undirected graph $\mathcal{G}_{T}=(\mathcal{V},\mathcal{E}_T)$ obtained by removing the orientation on all its edges, while avoiding repetition. An example of topology of the generative graph in Fig. \ref{fig:topexample}(a) is shown in Fig. \ref{fig:topexample}(b). The moral graph, $\mathcal{G}_{M}=(\mathcal{V},\mathcal{E}_M)$ of the generative graph $\mathcal{G}$, is defined as the undirected graph  obtained by removing the orientation on all its edges, avoiding repetition and adding an undirected edge between spouses. The moral graph of the generative graph in Fig. \ref{fig:topexample}(a) is shown in Fig. \ref{fig:topexample}(c).
Next, we present terminology which will be useful in the subsequent discussion.

\begin{definition}{\emph{Path:}}
A path between two nodes $x_0,x_k$ in an undirected graph $\mathcal{G}_T=(\mathcal{V},\mathcal{E}_T)$ is a set of unique nodes $\{x_0,x_1,\cdots, x_k\} \subseteq \mathcal{V}$ where $\{(x_0,x_1),\cdots,(x_{k-1},x_k)\} ~\subseteq \mathcal{E}_T$. We will denote a path by $x_0 - x_1 - x_2 -\cdots - x_{k-1} - x_k$. The length of a path is one less than the number of nodes in the path. For example: $1-2-3-5$ is a path of length three between node $1$ and $5$ in the undirected graph of \emph{Figure \ref{fig:topexample}(b)}.
\end{definition}

\begin{definition}{\emph{$m$ Hop Neighbor:}}
In an undirected graph $\mathcal{G}_T=(\mathcal{V},\mathcal{E}_T)$, $j \in \mathcal{V}$ is a $m$ hop neighbor of $i \in \mathcal{V}$, if there is a path of length $m$ between $i$ and $j$ in $\mathcal{G}_T$. For example: nodes $1$ and $5$ are three hop neighbors in the undirected graph in \emph{Figure \ref{fig:topexample}(b)}. If there is a path of length one between $i$ and $j$ then they are neighbors in $\mathcal{G}_T$. The set of $m$ hop neighbors of a node $i \in \mathcal{V}$ is denoted by $\mathcal{N}_{i,m}$.
\end{definition}

There are many systems which satisfy the dynamics represented by (\ref{eqn:lindyn}) some of which are discussed below.

\begin{enumerate}

 \item Consensus dynamics: Distributed decision making methods in multi-agent systems often use the first order consensus protocol, where each agent updates its states based on the difference between other agents' present value and itself, descibed by
  \begin{align}\label{eqn:consensus}
 \frac{dx_i}{dt} = \sum_{j =1}^{n}{c_{ij}}{(x_j-x_i)} + p_j,
 \end{align}
where, $p_j$ denotes the receiver noise for agent $j$ \cite{7827095}. Inferring, the communication topology of a network of agents in a multi agent system is a relevant objective of a cyber attacker, and appropriate hardware/ software tools need to be designed to avert such attacks. Here, based on the non zero values of $b_{ij}$, the generative graph $\mathcal{G}$ can be obtained.

\item RC networks: 
Tractable grey box modeling with lumped parameters prove effective for real-time control of buildings. Here $RC$ models that assume discretized physical space are employed where every identified zone is represented by a common temperature. Such a lumped parameter model is described by
\begin{align}\label{eqn:zone}
\mathcal{C}_i\frac{dT_i}{dt} = \sum_{j = 1}^{n}\frac{T_j-T_i}{R_{ij}} + p_j(t),
\end{align}
where, $\mathcal{C}_j>0$ is the capacitance of zone $j$, $R_{ij}\geq 0$ denotes the thermal resistance between node $i$ and $j$, ${p}_j$ is the total internal heat generated in zone ${j}$.
In the dynamics above, $R_{ji}=R_{ij}$ and thus in the description corresponding to (\ref{eqn:ldg}) with measured variable $x_i$ identified with temperature $T_i$, 
the underlying generative dynamics is bi-directional.
We consider the problem of inferring the influence topology using temperature sensors ubiquitously at possible zonal locations.

  \item Power Grid Network Dynamics: For small disturbances in the power grid, the dynamics of the deviation of voltage phase angle at bus $j$ from the nominal value, denoted by $\theta_j$, is modeled by the following linearized Swing Equation \cite{kundur1994power},
\begin{align}\label{swingmatrix}
M_i\ddot{\theta}_i + D_i\dot{\theta}_i = \sum_{j=1,\neq i}^{n} c_{ij}(\theta_j - \theta_i) + p_i(t),
\end{align}
where, $M_j$ denotes the inertia of the rotating mass, $D_j$ denotes the damping and $p_j(t)$ is the power imbalance injections at the node $j$. Here, $b_{ij}(\theta_j-\theta_i)$ gives the line flow from node $j$ to $i$, where $c_{ij}$ is the susceptance of the line. Under equilibrium conditions power balance is satisfied at each node. Here, $c_{ij} = c_{ji}$, implying, if $H_{ij}(z) \neq 0$ then $H_{ji}(z) \neq 0$ almost surely. Thus, similar to RC networks, the generative graph $\mathcal{G}$ has bi-directional edges. Inferring the topology of a power networks is often the first step to optimize flows and network monitoring for fault detection.
\end{enumerate}
Thermal dynamics and power grid dynamics are such that the physics of the system naturally lead to a bi-directed generative graph, where there is no clear notion of cause and effect among the nodes. Many other physical flow dynamics posses similar characteristics. Here, if $i,j$ are neighbors in the topology $\mathcal{G}_T$, then, $i$ is a parent as well as a child of $j$ in the generative graph. Furthermore, if $i,j$ are two hop neighbors in $\mathcal{G}_T$, then, $i$ and $j$ are spouses in the generative graph. The exact inference of the topology in these physical flow networks is equivalent to exact network inference of the bi-directed generative graph.

\section{Topology Learning}
In this article, the topology learning problem that we are interested in, involves inferring the underlying topology $\mathcal{G}_T$ of a linear dynamical system described by (\ref{eqn:lindyn}) with the generative graph $\mathcal{G}$, based solely on time series measurements $\{x_1(k),...,x_n(k)\}$ without the knowledge of the system parameters $\{a_{m,i}, b_{ij}\}$ where exogenous inputs $\{p_i\}$ are not measured.

 Earlier approaches considered the measurements as $iid$ (independent and identically distributed) samples of random variables with a joint distribution; here, states $\{x_j(k)\}_{j=1}^{n}$ are modeled as a collection of $iid$ samples of a multivariate Gaussian distribution. The framework is relevant when the sampling time $k$ is sufficiently far apart such that $\{x_j(k)\}_{j=1}^{n}$ satisfy the $iid$ sample requirement. Inference of the topology from $iid$ samples $\{x_j(k)\}_{j=1}^{n}$ (also known as Gaussian graphical model inference) is a well studied problem with Graph Lasso (maximum likelihood estimator of the topology from $iid$ samples) \cite{friedman2008sparse} being a well known approach. However, this framework becomes ineffective for high resolution data where dynamic effects between the measured variables are prominent. The inadequacies of the static framework of random variables is illustrated in  \cite{dekairep}, where, the authors show that these static approaches are unable to infer the true topology even with large data sets. Moreover, \cite{dekairep} shows that the static approach fails to correctly infer three node cycles present in power networks.

 Existing approaches in a dynamical setting with temporally correlated samples, model the exogenous inputs $p_i$  to be Gaussian white noise and independent from $p_j$ for $i\neq j$ \cite{pereira2010learning}. In this work we show that, one can infer the exact topology $\mathcal{G}_T$ of the linear dynamical system described by (\ref{eqn:lindyn}), even where exogenous inputs are colored, that is, correlated across time and can detect three node cycles accurately. We begin by recalling the idea of power spectral density.

\begin{definition}{\emph{Power Spectral Density(PSD) Matrix:}} For a $n$ dimensional collection of WSS time series $x(k)=\{x_1(k),...,x_n(k)\}^T$, the power spectral density matrix is defined as $\Phi_X(\omega) := \sum_{k=-\infty}^{\infty}\mathbb{E}(x(k)x(0)^T)e^{-\hat{j}\omega k}$, where, $\mathbb{E}(\cdot)$ denotes the expectation operator.
\end{definition}

\subsection{Multivariate Wiener Filtering}
Let $v$ and $x_1,...,x_m$ be a collection of jointly wide sense stationary (JWSS) stochastic processes. Let $x(k):=[x_{1}(k),....,x_{m}(k)]^{T}$ and $\mathcal{X}:=\textsl{span}\{x_1(k),...,x_m(k)\}_{k=-\infty}^{\infty}$. Consider the following least square optimization problem:
	\begin{align}
		\hat{v}(k) := \arg\inf_{q \in \mathcal{X}} \mathbb{E}(v(k)-q)^2. \label{wiener}
	\end{align}
	If $\Phi_{X}(\omega)\succ 0$ (that is $\Phi_{X}(\omega)$ is positive definite) almost surely, then the optimal solution $\hat{v}(k) \in \mathcal{X}$ exists, is unique and is given by
	\begin{align*}
		\hat{V}(z) &= \textbf{W}(z)X(z), \textbf{W}(z)=\Phi_{v X}(z)\Phi_{X}(z)^{-1}\\
		&= [W_{v1}(z) \ ... \ W_{vm}(z)]X(z)\\
		&= \sum_{j=1}^{m} W_{vj}(z)X_j(z),
	\end{align*}
where $\textbf{W}(z)=\Phi_{v X}(z)\Phi_{X}(z)^{-1}$ is the Wiener filter. Here, $\hat{V}(z)$ is the $z$ transform of $\hat{v}(k)$. Refer \cite{materassi2012problem} for further details.
The next result details the properties of Wiener filter for topology inference of linear dynamical systems described by (\ref{eqn:lindyn}).

\subsection{Learning the Moral Graph using Multivariate Wiener Filtering}
\begin{thm}\label{thm:sparse_wiener}
	Consider the linear dynamical system in \emph{(\ref{eqn:lindyn})} with topology $\mathcal{G}_T$ and $({H}(z),{E}(z))$ specifying the network dynamics in \emph{(\ref{eqn:netdyn})}. The nodal state measurements are given by ${x}(k)=[{x}_{1}(k),...,{x}_{n}(k)]^{T}$.
	Define the space $\mathcal{X}_{\bar{j}}=\textsl{span}\{{x}_i(k)\}_{i\neq j,k=-\infty}^{k=\infty}$.
	The non-causal Multivariate Wiener filter estimate ${\hat{{x}}}_j(k)\in \mathcal{X}_{\bar{j}}$ of the signal ${x}_j(k)$ is given by,
	\begin{align}
		\hat{{X}}_j(z) =\sum_{i\neq j,i=1}^{n} {W}_{ji}(z){X}_i(z),
	\end{align}
	where, ${W}_{ji}(z)\neq 0$ implies $i \in \mathcal{N}_j \cup K_{j}$ in $\mathcal{G}_T$.
\end{thm}
\begin{pf}
The proof follows from the main result of \cite{materassi2012problem}, which states that, $W_{ji}(z) \neq 0$ implies that $j$ and $i$ share a children-parent or spouse relationship in the generative graph $\mathcal{G}$. It follows that, in terms of the topology $\mathcal{G}_T$, if $W_{ji}(z)\neq 0$ then $i$ and $j$ are neighbors or spouses.
\end{pf}
\begin{remark}
 The above result does not guarantee that if $i \in \mathcal{N}_j \cup {K}_{j}$, then $W_{ji}(z) \neq 0$. However, such cases are pathological \emph{(see {\cite{materassi2012problem}})}.
\end{remark}

Thus the set of children, parents and spouses of each node in the network can be identified using non-zero entries in the corresponding multivariate Wiener filter. The moral graph, $\mathcal{G}_M$, can be obtained by adding a link between nodes with non zero entries in the corresponding multivariate Wiener filter. We summarize the procedure to obtain $\mathcal{G}_M$ from nodal time series measurement in Algorithm $1$ below.
\begin{algorithm}
\caption{Learning Moral Graph using Wiener Filtering}
\textbf{Input:} samples $x_i(k)$ for nodes $i \in \{1,2,...,n\}$ from generative graph $\mathcal{G}$, thresholds $\rho,\tau$, frequency points $ \Omega$ = \{$\omega_1,...,\omega_m $\} where $\omega_i \in [-\pi,\pi]$\\
\textbf{Output:} Estimate of edges of $\mathcal{G}_M$, ${\bar{\mathcal{E}}}_M$(for large number of samples per node, ${\bar{\mathcal{E}}}_M$ coincides with ${\mathcal{E}}_M$)\\
\begin{algorithmic}[1]
\State Edge set ${\bar{\mathcal{E}}_M} \gets \{\}$
\ForAll{$j \in \{1,2,...,n\}$}\label{step1_a}
\State Compute Wiener filter ${W}_j(e^{\hat{j}\omega}) = [W_{j1}(e^{\hat{j}\omega})\ \cdots W_{j,j-1}(e^{\hat{j}\omega}), W_{j,j+1}(e^{\hat{j}\omega}) \cdots W_{jn}(e^{\hat{j}\omega})]$ \label{step1_a1}
\EndFor

\ForAll{$i,j \in \{1,2,...,n\}, i\neq j$}
\If{$ \sup_{\omega_i \in \Omega}  \|W_{ji}(e^{\hat{j}\omega_i})\|>\rho$}\label{step_nonzero}
\State $\bar{\mathcal{E}}_M \gets \bar{\mathcal{E}}_M \cup \{(i,j)\}$
\EndIf
\EndFor\label{step1_b}
\end{algorithmic}
\end{algorithm}

\begin{figure}[tb]
	\centering
	\begin{tabular}{ccc}
		\includegraphics[width=0.8\columnwidth]{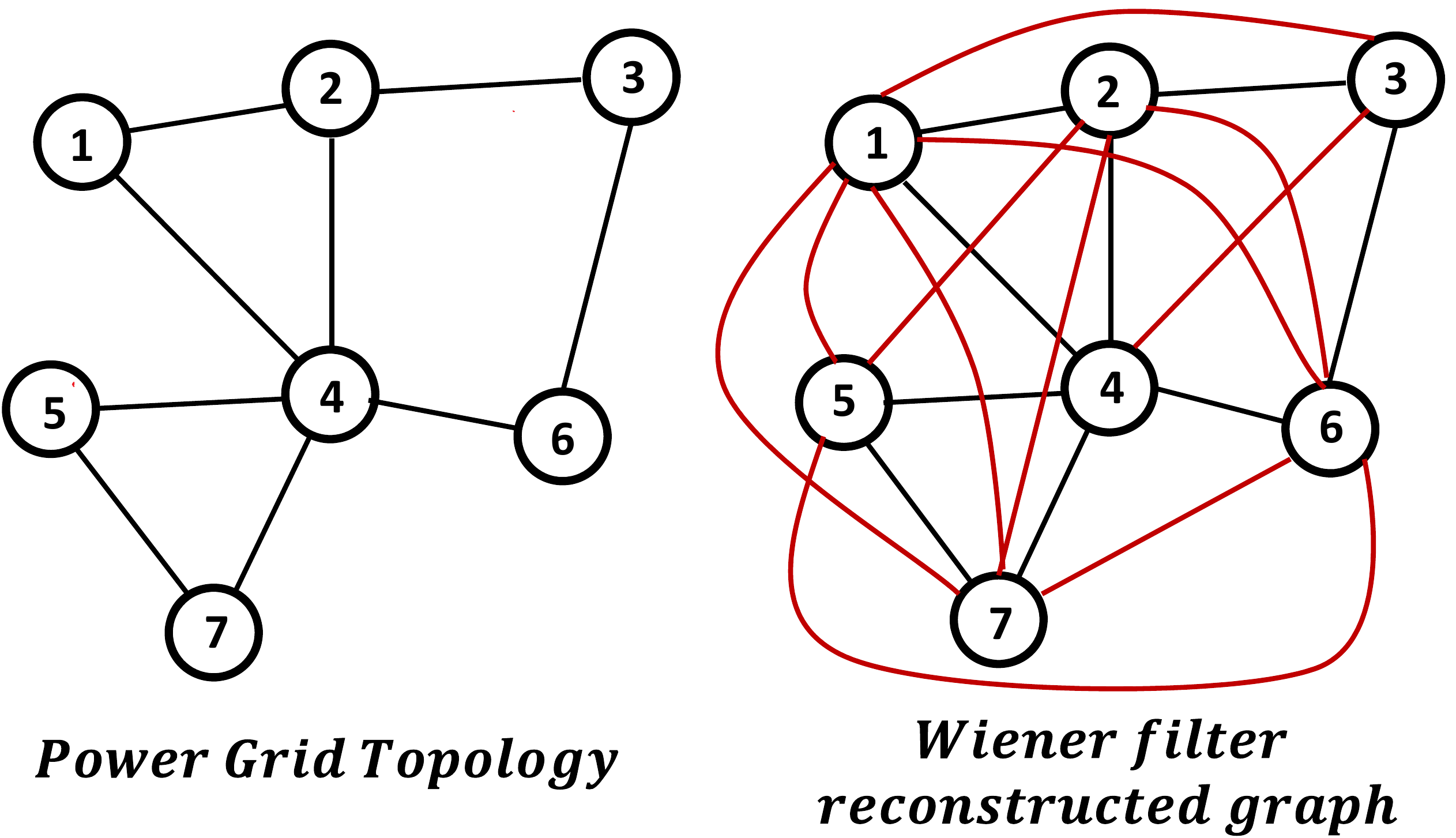}\\
		(a)
	\end{tabular}
	\begin{tabular}{ccc}
		\includegraphics[width=0.8\columnwidth]{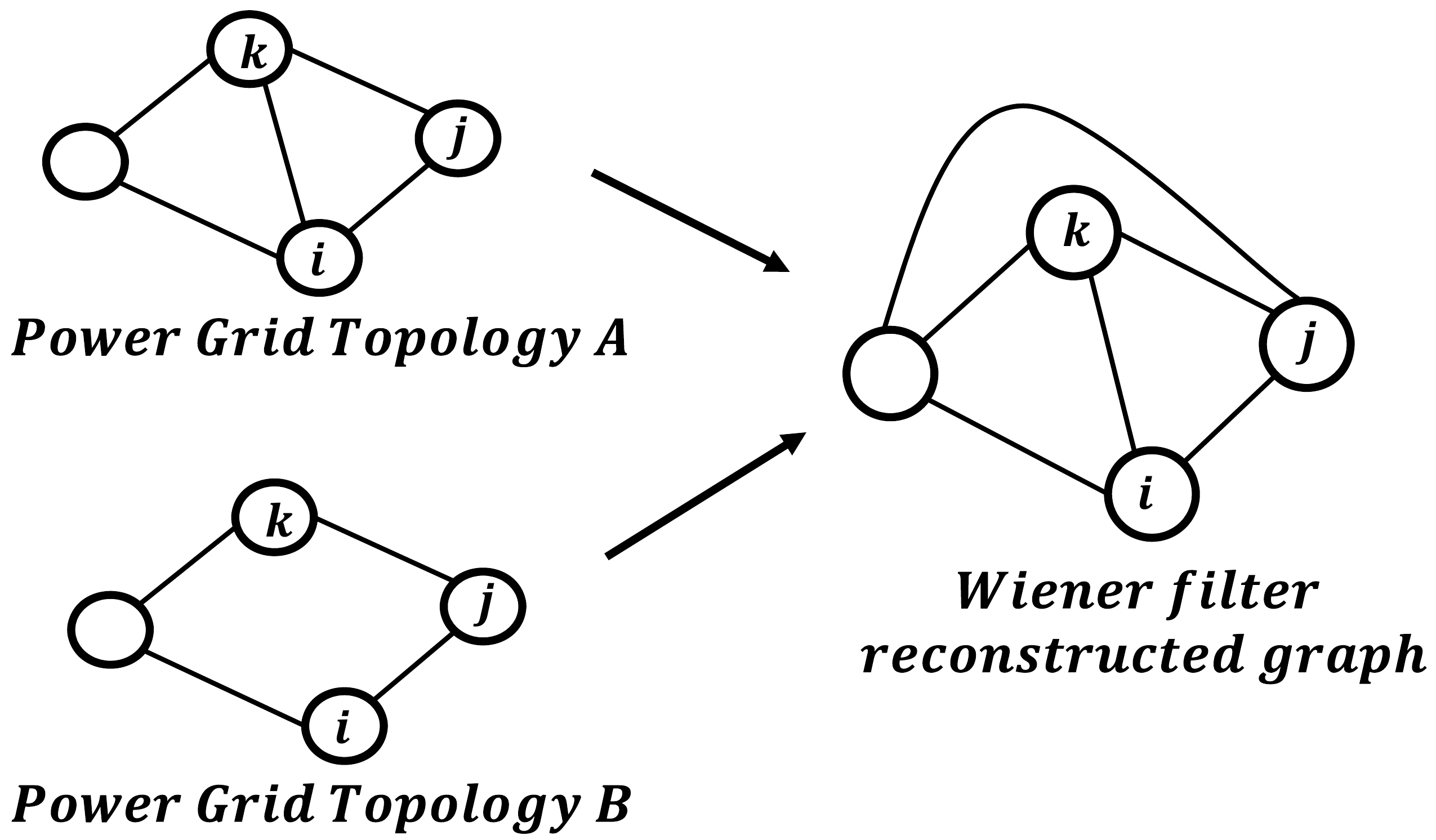}\\
		(b)
	\end{tabular}
	\squeezeup
	\caption{\small{(a) Number of spurious links (red edges) in the Wiener reconstructed graph are comparable with the number of true links (black edges) in the underlying power grid topology (b) Example of two power grid topology which result in the same reconstructed graph of non-trivial Wiener filters}}
		\label{fig:example}
\end{figure}

Algorithm $1$ results in the moral graph, which has spurious spouse edges apart from the edges present in the topology. For bi-directed generative graphs considered here, the number of spurious links in the moral graph is of the same order as that of true links; here, for every pair of two hop neighbor we get one edge in the moral graph which is not present in the generative graph. Indeed, consider the power grid topology in Fig. \ref{fig:example}(a) and its moral graph; it is evident that the number of spurious links (links not present in the topology) is substantial. Furthermore, consider the example in Fig.~\ref{fig:example} (b), where both the topology $A$ and $B$ obtained from the generative graphs of two power grids, following the swing equation dynamics, result in the same reconstructed moral graph using multivariate Wiener filtering. The Wiener filtering based reconstructed edge between $i$ and $k$ in Fig.~\ref{fig:example} can thus represent a true edge or a spurious edge between spouses. Thus, elimination of spurious edges to recover the actual topology is a non trivial and important task owing to the significant presence of spurious links and presence of many candidate topologies for a given moral graph.

For radial topologies (undirected connected graph with no cycles) associated with bi-directed generative graphs, it is possible to distinguish between true edges between neighbors and spurious two hop neighbor edges in $\mathcal{G}_M$ by using a local graph separation rule as presented in \cite{talukdar2017exact}. However for bi-directed generative graphs with topologies having cycles or loops, such graph separation results do not hold in general \cite{dekairep}. In the next section, we present methods that eliminate spurious links obtained from Algorithm $1$, for a perfect reconstruction of the generative graph. Here the physics of the dynamics given by (\ref{eqn:lindyn}) will prove crucial.

\section{Exact Reconstruction of the generative graph}\label{sec:pruning}
The following theorem presents an explicit characterization of the contribution of neighbors $\mathcal{N}_j$ and two-hop neighbors $\mathcal{N}_{j,2}$ of node $j$ to the multivariate non causal Wiener filter $\{{W}_{ji}(z)\}_{i=1,i \neq j}^{n}$.

\begin{thm}\label{thm:expression}
Consider the generative graph $\mathcal{G} = (\mathcal{V}, \mathcal{E})$ described by \emph{(\ref{eqn:lindyn})}, with $x(k)=(x_1(k) \ \cdots \ x_n(k))^{T}$ as the output at time instant $k$. Let the $z$ transform of the multivariate non causal Wiener filtering estimate $\hat{x}_j(k)$ of $x_j(k)$ be,
$\hat{X}_j(z)= \sum_{i,i\neq j}$ $W_{ji}(z)X_i(z)$. Then, $W_{ji}(z) = \hat{C}_{ji}(z) + \hat{P}_{ji}(z) + \hat{K}_{ji}(z) $, where,
\squeezeup
\begin{align}
\hat{C}_{ji}(z) &= \frac{b_{ij}S_i(z)\Phi_{p_i}^{-1}(z)}{|S_j(z)|^2\Phi_{p_j}^{-1}(z)+\sum_{l\in P_j}b_{lj}^2\Phi_{p_l}^{-1}(z)}\label{eqn:childfactors},\\
\hat{P}_{ji}(z) &= \frac{b_{ji}S_j^{*}(z)\Phi_{p_j}^{-1}(z)}{|S_j(z)|^2\Phi_{p_j}^{-1}(z) + \sum_{l\in \mathcal{N}_j}b_{lj}^2\Phi_{p_l}^{-1}(z)}\label{eqn:parentfactors},\ \text{and},\\
\hat{K}_{ji}(z) &=\displaystyle -\frac{\sum_{k\in \mathcal{N}_i\cap \mathcal{N}_j}b_{kj}b_{ki}\Phi_{p_k}^{-1}(z) }{|S_j(z)|^2\Phi_{p_j}^{-1}(z) + \sum_{l\in P_j}b_{lj}^2\Phi_{p_l}^{-1}(z)}. \label{eqn:spousefactors}
\end{align}
\end{thm}
\squeezeup
\begin{pf}
It is shown in \cite{materassi2012problem} that,
$W_{ji}(z) = \hat{C}_{ji}(z) + \hat{P}_{ji}(z) + \hat{K}_{ji}(z)$, where,
\squeezeup
\begin{align*}
\hat{C}_{j*}(z) &= \frac{\Phi_{e_j}(z)H_{*j}^{*}(z)\Phi_{e}^{-1}(z)}{1+|H_{*j}^{*}(z)\Phi_e^{-1}H_{*j}(z)|\Phi_{e_j}(z)},\\
\hat{P}_{j*}(z) &=(1-\hat{C}_{j*}(z)H_{*j}(z))H_{j*}(z),\\
\hat{K}_{ji}(z) &= -\hat{C}_{j*}(z)H_{*i}(z),
\end{align*}
with $H_{*j}(z)$ and $H_{j*}(z)$ representing the $j-th$ column and $j-th$ row of the $H(z)$ matrix respectively and $H_{*j}^*(z)$ represents complex conjugate transpose of the vector $H_{*j}(z)$.  Here $\hat{C}_{ji}$, $\hat{P}_{ji}$ and $\hat{K}_{ji}$ are contributions of the $i^{th}$ time-series in the estimation of the $j^{th}$ time-series for being the child, parent  and spouse of $j$ respectively. Note that $\hat{C}_{jj}(z) = 0$ and $\hat{P}_{jj}(z) = 0$.
In the context of a the generative graph $\mathcal{G}$ with dynamics described by (\ref{eqn:lindyn}), $\hat{C}_{j*}(z)$ is a $1 \times n$ row of transfer functions described by,
\begin{align*}
\hat{C}_{j*}(z) &= \frac{\Phi_{e_j}(z)H_{*j}^{*}(z)\Phi_e^{-1}(z)}{1+|H_{*j}^{*}(z)\Phi_{e}^{-1}H_{*j}(z)|\Phi_{e_j}(z)}\\
&= \frac{[H_{1j}^{*}(z)\Phi_{e_1}^{-1}(z) \ H_{2j}^{*}(z)\Phi_{e_2}^{-1}(z) \ \cdots \ H_{nj}^{*}(z)\Phi_{e_n}^{-1}(z)]}{\Phi_{e_j}^{-1}(z)+\sum_{l\in P_j}|H_{lj}(z)|^2\Phi_{e_l}^{-1}(z)}\\
&= \frac{[\frac{b_{1j}}{S_1^{*}(z)}|S_1(z)|^2\Phi_{p_1}^{-1}(z) \ \cdots \ \frac{b_{nj}}{S_n^{*}(z)}|S_n(z)|^2\Phi_{p_n}^{-1}(z)]}{|S_j(z)|^2\Phi_{p_j}^{-1}(z)+\sum_{l\in P_j}\frac{b_{lj}^2}{|S_{l}(z)|^2}|S_{l}(z)|^{2}\Phi_{p_l}^{-1}(z)}\\
&= \frac{[{b_{1j}}S_1(z)\Phi_{p_1}^{-1}(z) \ \cdots \ {b_{nj}}S_n(z)\Phi_{p_n}^{-1}(z)]}{|S_j(z)|^2\Phi_{p_j}^{-1}(z)+\sum_{l\in P_j}{b_{lj}^2}\Phi_{p_l}^{-1}(z)}.
\end{align*}
Thus, contribution to $W_{ji}(z)$ due to $i$ being a child of $j$ is given by, $\hat{C}_{ji}(z) = \frac{b_{ij}S_i(z)\Phi_{p_i}^{-1}(z)}{|S_j(z)|^2\Phi_{p_j}^{-1}(z)+\sum_{l\in P_j}{b_{lj}^2}\Phi_{p_l}^{-1}(z)}$. Similarly, contribution from all parents of node $j$ in the generative graph $\mathcal{G}$, $\hat{P}_{j*}(z)$ is a $1 \times n$ row of transfer functions described by,
\begin{align*}
&\hat{P}_{j*}(z) = (1 - \hat{C}_{j*}(z)H_{*j}(z))H_{j*}(z)\\
&=(1 - \frac{\sum_{l \in N_j}{b_{lj}^2}\Phi_{p_l}^{-1}(z) }{|S_j(z)|^2\Phi_{p_j}^{-1}(z)+\sum_{l\in P_j}{b_{lj}^2}\Phi_{p_l}^{-1}(z)})\\
&[H_{j1}(z)\ \cdots \ H_{jn}(z)]\\
&= \frac{|S_j(z)|^2\Phi_{p_j}^{-1}(z)}{|S_j(z)|^2\Phi_{p_j}^{-1}(z)+\sum_{l\in P_j}{b_{lj}^2}\Phi_{p_l}^{-1}(z)}\frac{1}{S_j(z)}\\
&[b_{j1}(z)\ \cdots \ b_{jn}(z)]\\
&=\frac{S_j^{*}(z)\Phi_{p_j}^{-1}(z)}{|S_j(z)|^2\Phi_{p_j}^{-1}(z)+\sum_{l\in P_j}{b_{lj}^2}\Phi_{p_l}^{-1}(z)}[b_{j1}(z)\ \cdots \ b_{jn}(z)].
\end{align*}
Thus, the contribution to $W_{ji}(z)$ due to $i$ being a parent of $j$ is $\hat{P}_{ji}(z) = \frac{b_{ji}S_j^*(z)\Phi_{p_j}^{-1}(z)}{|S_j(z)|^2\Phi_{p_j}^{-1}(z)+\sum_{l\in P_j}{b_{lj}^2}\Phi_{p_l}^{-1}(z)}$. The net contribution to $W_{ji}(z)$ due to $i \in \mathcal{N}_j$, is given by,
$\hat{N}_{ji}(z) = \hat{C}_{ji}(z) + \hat{P}_{ji}(z)$. The contribution, $\hat{K}_{ji}$,  to $W_{ji}$ for $i$ for  being a spouse of $j$ in the generative graph $\mathcal{G}$ with dynamics described by (\ref{eqn:lindyn}) is,
\begin{align*}
\hat{K}_{ji}(z)&= -\hat{C}_{j*}(z)H_{*i}\\
&=-\displaystyle \frac{\sum_{k\in C_j \cap C_i}b_{kj}S_k(z)\frac{b_{ki}}{S_k(z)}\Phi_{p_k}^{-1}(z)}{|S_j(z)|^2\Phi_{p_j}^{-1}(z)+\sum_{l\in P_j}{b_{lj}^2}\Phi_{p_l}^{-1}(z)}\\
&=-\displaystyle \frac{\sum_{k\in C_j \cap C_i}b_{kj}{b_{ki}}\Phi_{p_k}^{-1}(z)}{|S_j(z)|^2\Phi_{p_j}^{-1}(z)+\sum_{l\in P_j}{b_{lj}^2}\Phi_{p_l}^{-1}(z)}.
\end{align*}
This proves the theorem.
\end{pf}
Next, we use the expressions derived in Theorem \ref{thm:expression} to distinguish between true and spurious edges in the moral graph $\mathcal{G}_M$ formed by non-zero entries of multivariate Wiener filter. The next theorem presents a result using the phase response of the non causal Wiener filter for spurious edges corresponding to strict spouse relationships, enabling them to be distinguished from true edges.
\begin{thm}\label{thm:piresult}
Consider the generative graph $\mathcal{G} = (\mathcal{V}, \mathcal{E})$ described by \emph{(\ref{eqn:lindyn})}, with $x(k)=(x_1(k) \ \cdots \ x_n(k))^{T}$ as the output at time instant $k$. If $i$ and $j$ are strict spouses in $\mathcal{G}$, that is, ${C}_i \cap {C}_j \neq \phi$ and $i \notin \mathcal{N}_j$, $j \notin \mathcal{N}_i$, then $\angle (W_{ji}(e^{\hat{j}\omega})) = \pi$ for all $\omega \in [-\pi,\pi]$.
\end{thm}
\begin{pf}
As $i \notin \mathcal{N}_j$ and $j \notin \mathcal{N}_i$, $b_{ij} = 0, b_{ji} = 0$. It follows from Theorem \ref{thm:expression} that, $\hat{N}_{ji}(z) = 0$. Since, $i$ and $j$ are spouses, there exist $k \in C_i \cap C_j$ such that $b_{kj} > 0$ and $b_{ki} > 0$, implying $\hat{K}_{ji}(z) \neq 0$. Since, $\hat{K}_{ji}(z)$ is dependent on $\{\Phi_{p_j}(z)\}_{j=1,2,\cdots,n}$, which are positive real numbers for WSS processes, it follows that, $\hat{K}_{ji}(z) < 0$ for all $z \in \mathbb{C}$. Thus,
\begin{align*}
\angle (W_{ji}(e^{\hat{j}\omega})) =\angle (\hat{K}_{ji}(e^{\hat{j}\omega}))= \pi, \text{for all}\ \omega \in [-\pi,\pi].
\end{align*}
\end{pf}
We will use the above theorem to identify spurious edges in the moral graph $\mathcal{G}_M$ to recover the true topology $\mathcal{G}_T$. We now show that the above result does not hold for nodes that are neighbors. The next theorem lists a condition under which $\angle (W_{ji}(e^{\hat{j}\omega})) = \pi$ for all $\omega \in [-\pi,\pi]$ for nodes $i$ and $j$ that are neighbors as well as spouses.
\begin{thm}\label{thm:notpiresult1}
Consider the generative graph $\mathcal{G} = (\mathcal{V}, \mathcal{E})$ described by \emph{(\ref{eqn:lindyn})}, with $x(k)=(x_1(k) \ \cdots \ x_n(k))^{T}$ as the output at time instant $k$. Suppose $i$ and $j$ are such that $i \in \mathcal{N}_j$, $i \in {K}_{j}$. Then $\angle (W_{ji}(e^{\hat{j}\omega})) = \pi$ for all $\omega \in [-\pi,\pi]$, if and only if,
\begin{align*}
 &Im(b_{ij}S_i(e^{\hat{j}\omega})\Phi_{p_i}^{-1}(e^{\hat{j}\omega})+
 b_{ji}S_j^{*}(e^{\hat{j}\omega})\Phi_{p_j}^{-1}(e^{\hat{j}\omega}))=0, \\
 &Re(b_{ij}S_i(e^{\hat{j}\omega})\Phi_{p_i}^{-1}(e^{\hat{j}\omega})+b_{ji}S_j^{*}(e^{\hat{j}\omega})\Phi_{p_j}^{-1}(e^{\hat{j}\omega})) -\\
 &\sum_{k\in \mathcal{N}_i\cap \mathcal{N}_j}b_{kj}b_{ki}\Phi_{p_k}^{-1}(e^{\hat{j}\omega})<0.
\end{align*}
for all $\omega \in [-\pi,\pi]$, where $Im(z)$ and $Re(z)$ represent imaginary and real parts of the complex number $z$.
\end{thm}

\begin{pf}
Using (\ref{eqn:childfactors}), (\ref{eqn:parentfactors}) and (\ref{eqn:spousefactors}),
\begin{align}
 W_{ji}(e^{\hat{j}\omega}) &= \frac{b_{ij}S_i(e^{\hat{j}\omega})\Phi_{p_i}^{-1}(e^{\hat{j}\omega})+b_{ji}S_j^{*}(e^{\hat{j}\omega})\Phi_{p_j}^{-1}(e^{\hat{j}\omega}) }{|S_j(e^{\hat{j}\omega})|^2\Phi_{p_j}^{-1}(e^{\hat{j}\omega}) + \sum_{l\in P_j}b_{lj}^2\Phi_{p_l}^{-1}(e^{\hat{j}\omega})}\nonumber\\
 &-\frac{\sum_{k\in \mathcal{N}_i\cap \mathcal{N}_j}b_{kj}b_{ki}\Phi_{p_k}^{-1}(e^{\hat{j}\omega})}{|S_j(e^{\hat{j}\omega})|^2\Phi_{p_j}^{-1}(e^{\hat{j}\omega}) + \sum_{l\in P_j}b_{lj}^2\Phi_{p_l}^{-1}(e^{\hat{j}\omega})}\label{eqn:Wji}
\end{align}
The second term in the above expression has no imaginary component and the denominator of both the terms are real.

$(\Rightarrow)$
Since, $\angle (W_{ji}(e^{\hat{j}\omega})) = \pi$ for all $\omega \in [-\pi,\pi]$ it follows that the conditions hypothesized in the theorem statement hold.

$(\Leftarrow)$
If the system parameters satisfy the conditions described in the only if part of the theorem statement, it follows from (\ref{eqn:Wji}) that,
$\angle (W_{ji}(e^{\hat{j}\omega})) = \pi$ for all $\omega \in [-\pi,\pi]$.
\end{pf}

\begin{remark}
The conditions presented in the previous theorem are such that for neighbor nodes $i$ and $j$ which are also spouses, $\angle(W_{ji}(e^{\hat{j}\omega})) = \pi$ for all $\omega \in [-\pi,\pi]$, is pathological because the system parameters have to take a specific set of values for the above mentioned conditions to be true at all frequencies and hence is pathological.
\end{remark}
Finally, the next theorem shows that $\angle (W_{ji}(e^{\hat{j}\omega})) = 0$ for $\omega = 0$ when $i$ and $j$ are neighbors and not spouses.

\begin{thm}\label{thm:notpiresult2}
Consider the generative graph $\mathcal{G} = (\mathcal{V}, \mathcal{E})$ described by \emph{(\ref{eqn:lindyn})}, with $X(k)=(x_1(k) \ \cdots \ x_n(k))^{T}$ as the output at time instant $k$. Nodes $i$ and $j$ are such that $i \in \mathcal{N}_j$ and $i\not\in K_{j}$. Then, $\angle (W_{ji}(e^{\hat{j}\omega}))|_{\omega = 0} = 0 $.
\end{thm}

\begin{pf}
Since $i \not\in K_{j}$, $\hat{K}_{ji}(z)=0$. It follows that,
\begin{align}\label{eqn:Wji2}
 W_{ji}(e^{\hat{j}\omega}) &= \hat{N}_{ji}(z)\nonumber \\
 &= \frac{b_{ij}S_i(e^{\hat{j}\omega})\Phi_{p_i}^{-1}(e^{\hat{j}\omega})+b_{ji}S_j^{*}(e^{\hat{j}\omega})\Phi_{p_j}^{-1}(e^{\hat{j}\omega}) }{|S_j(e^{\hat{j}\omega})|^2\Phi_{p_j}^{-1}(e^{\hat{j}\omega}) + \sum_{l\in \mathcal{N}_j}b_{lj}^2\Phi_{p_l}^{-1}(e^{\hat{j}\omega})}.
\end{align}
The denominator of the expression on the right hand side of (\ref{eqn:Wji2}) is real and positive for all $\omega \in [-\pi,\pi]$. The numerator of the expression on the right hand side of (\ref{eqn:Wji2}), $[b_{ij}S_i(e^{\hat{j}\omega})\Phi_{p_i}^{-1}(e^{\hat{j}\omega})+b_{ij}S_j^{*}(e^{\hat{j}\omega})\Phi_{p_j}^{-1}(e^{\hat{j}\omega})]_{\omega = 0}$ is also real and positive. Thus, $ W_{ji}(e^{\hat{j}\omega})|_{\omega=0}$ is real and positive, implying, $\angle (W_{ji}(e^{\hat{j}\omega}))|_{\omega= 0} = 0$.
\end{pf}

\begin{remark}
 Theorems \ref{thm:piresult}, \ref{thm:notpiresult1} and \ref{thm:notpiresult2} demonstrate that aside from pathological cases, the phase of the Wiener filter $\angle (W_{ji}(e^{\hat{j}\omega})) = \pi$ for all $\omega \in [-\pi,\pi]$ only when $i$ and $j$ are \textbf{not} neighbors but are spouses in the generative graph $\mathcal{G}$. In other words, aside for the pathological cases, the converse of Theorem \ref{thm:piresult} holds and can be used as a criteria to differentiate between true edges and spurious edges in the moral graph $\mathcal{G}_M$.
\end{remark}

We now present Algorithm $2$ that estimates the topology of the generative graph $\mathcal{G}$ based on time-series of nodal measurements pertaining to dynamics expressed by (\ref{eqn:lindyn}). The algorithm consists of two parts. The first part (Steps \ref{step1_a} - \ref{step1_b}) determines the multivariate Wiener filter $W_{ji}(z)$ to estimate the moral graph and is same as Algorithm $1$. In the next part (Steps \ref{step2_a} - \ref{step2_b}), we consider a representative set of frequency points $\Omega$ in the interval $[-\pi,\pi)$ and evaluate the phase angle of the Wiener filters for edges in ${\mathcal{E}}^{w}$. If the phase angle is within a pre-defined threshold $\tau$ of $-\pi$, the algorithm designates them as spurious edges (see Theorem \ref{thm:piresult}) and prunes them from ${\mathcal{E}}^{w}$ to produce edge set $\bar{{\mathcal{E}}}$ of the estimated true topology.

\begin{algorithm}
\caption{Topology Learning using Wiener Filtering with Pruning Step}
\textbf{Input:} nodal time samples $x_i(k)$ for nodes $i \in \{1,2,...,n\}$ in the generative graph $\mathcal{G}$, thresholds $\rho, \tau$, frequency points $ \Omega$ = \{$\omega_1,...,\omega_m $\} where $\omega_i \in [-\pi,\pi]$\\
\textbf{Output:} Estimate of Edges $\bar{{\mathcal{E}}}_T$ in the topology of $\mathcal{G}$. For large samples, $\bar{{\mathcal{E}}}_T$ coincides with ${\mathcal{E}}_T$\\
\begin{algorithmic}[1]
\State Edge set ${\mathcal{E}}^{w} \gets \{\}$
\ForAll{$j \in \{1,2,...,n\}$}\label{step1_a}
\State Compute Wiener filter ${W}_j(e^{\hat{j}\omega}) = [W_{j1}(e^{\hat{j}\omega})\ \cdots W_{jn}(e^{\hat{j}\omega})]$ \label{step1_a1}
\EndFor
\ForAll{$i,j \in \{1,2,...,n\}, i\neq j$}
\If{$ \sup_{\omega_i \in \Omega} \|W_{ji}(e^{\hat{j}\omega_i})\| > \rho$}\label{step_nonzero}
\State $\mathcal{E}^w \gets \mathcal{E}^w \cup \{(i,j)\}$
\EndIf
\EndFor\label{step1_b}
\State Edge set $\bar{\mathcal{E}}_T \gets {\mathcal{E}}^{w}$ \label{step2_a}
\ForAll{$i,j \in \{1,2,...,n\}, i\neq j$}
\If{$\pi -\tau \leq |\angle(W_{ji}(e^{\hat{j}\omega_i}))| \leq \pi +\tau, \forall \omega_i \in \Omega$}
\State $\bar{\mathcal{E}}_T \gets \bar{\mathcal{E}}_T - \{(i,j)\}$
\EndIf
\EndFor \label{step2_b}
\end{algorithmic}
\end{algorithm}

\begin{figure}[tb]
	\centering
		\includegraphics[scale=0.35]{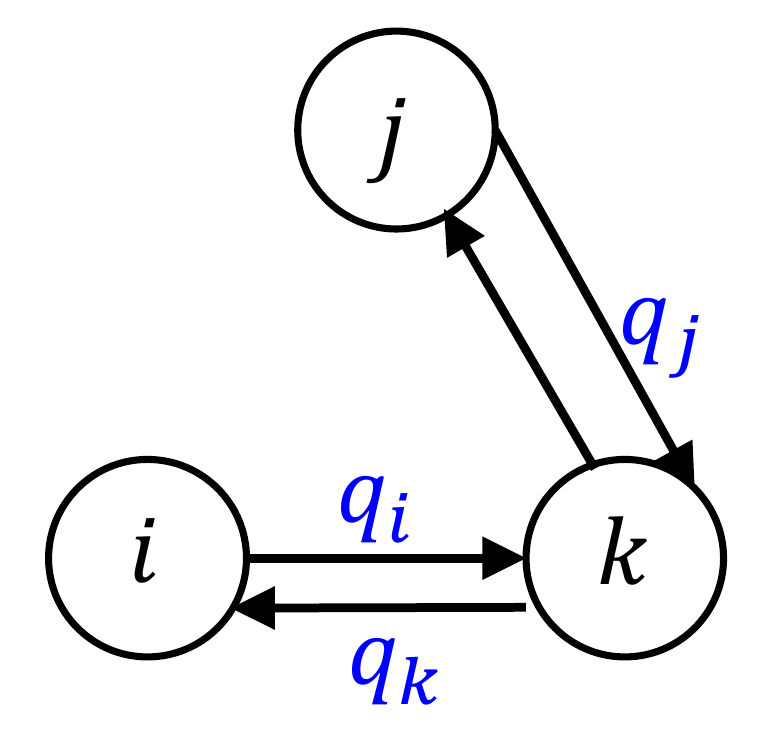}
	\caption{A three node system, with the flow on the edge indicated in blue.\textcolor{red}{$q_k$ needs to be reassigned in the figure right now it looks like going to i}}
		\label{fig:flowcons}
\end{figure}

\textbf{Interpretation of Pruning Step for Physical Flow Networks}
Consider nodes $i,j,k$ in a physical flow system like a thermal RC network or a power network, where $k$ is a common neighbor of both $i,j$ but $i,j$ are not neighbors. Let $q_i,q_j$ denote the flow out of node $i,j$ respectively while $q_k$ is the total flow received at node $k$ as shown in Figure \ref{fig:flowcons}. Here, $q_i,q_j,q_k$ could represent flow of heat, electrical power or even a fluid driven by the difference in temperature, voltage or pressure respectively. These flow variables $q_i,q_j,q_k$ are directly proportional to the nodal state $x_i,x_j,x_k$ respectively.  It follows from flow conservation,
\begin{align*}
  q_k &= q_i + q_j,\\
  q_i&=q_k - q_j,
\end{align*}
where, a negative correlation is observed between $q_i$ and $q_j$ owing to flow conservation constraint. This translates into an inverse relationship  between nodal state variables $x_i$ and $x_j$ leading to phase $\pi$ relationship of the corresponding Wiener filter. Thus, the flow conservation in physical flow networks translates into phase angle being $\pi$ for the associated Wiener filters providing a physics based pruning step in our topology learning algorithm.

\section{Simulation and Experimental Validation}

In this section we present simulation and experimental validation of the topology inference algorithm presented in the previous section. First, we present the method employed  to compute the Wiener filter using nodal time series measurements and is used in Algorithm $2$.

\textbf{Wiener filter computation}
Let $x_j(k)$ be the nodal time series to be estimated from $\{x_i(k)\}_{i=1,i\neq j}^{m}, k\in \mathbb{Z}$. Consider the following least square optimization problem on the Hilbert space of $\mathcal{L}_2$ random variables,
\begin{align} \label{wiener2}
\{h_{ji,o}\}&={\arg \inf_{{\{h_{ji}\}}_{i=1,...,m,i\neq j}}}~~ \nonumber\\
&\mathbb{E}(x_j(k)- \sum_{i=1,i\neq j}^{m}\sum_{L=-F}^{F}h_{ji}^{L}x_i(k-L))^2,
\end{align}
where, $h_{ji}=[h_{ji}^{-F},...,h_{ji}^{0},...,h_{ji}^{F}]$.
From the implementation viewpoint, we consider lags up to a finite order $F$ in (\ref{wiener}).	The solution to the above optimization problem is referred as the \emph{finite impulse response multivariate non-causal Wiener filter}) \cite{huang2006acoustic}, $W_j(z)=[W_{j,1}(z),...,W_{j,j-1}(z),W_{j,j+1}(z),...,W_{j,m}(z)]$, where,
\begin{align}\label{solform}
	  W_{ji}(z) = \sum_{L=-F}^{F}h_{ji,o}^{L}z^{-L}.
	\end{align}
It is important to note that the multivariate Wiener filter, which is obtained from solving the above optimization problem is determined entirely from the measured time series without any knowledge of system parameters or the statistics of the exogenous inputs. Next, we demonstrate the effectiveness of Algorithm $2$ for inference of network structure of agents undergoing linear consensus dynamics.

\subsection{Validation on Consensus Dynamics}

\begin{figure}[tb]
	\centering
	\begin{tabular}{ccc}
	\includegraphics[width=0.3\columnwidth]{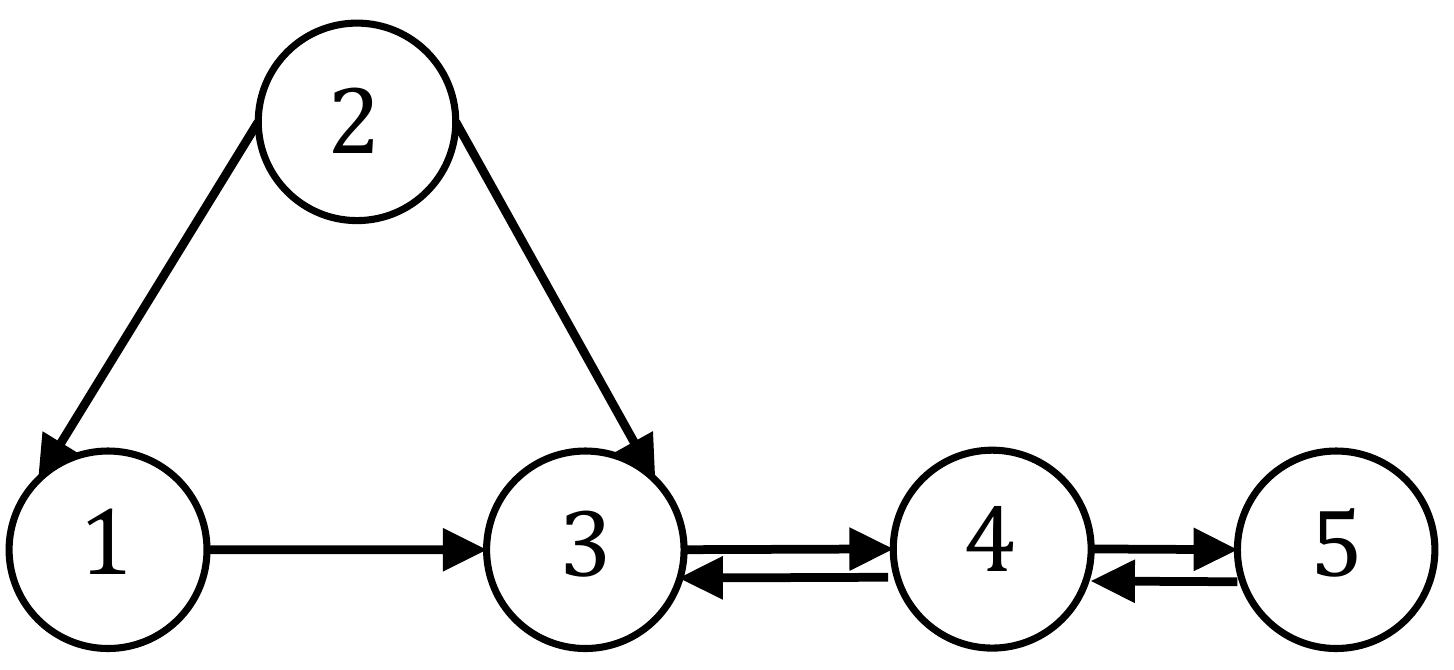}
	&
		\includegraphics[width=0.3\columnwidth]{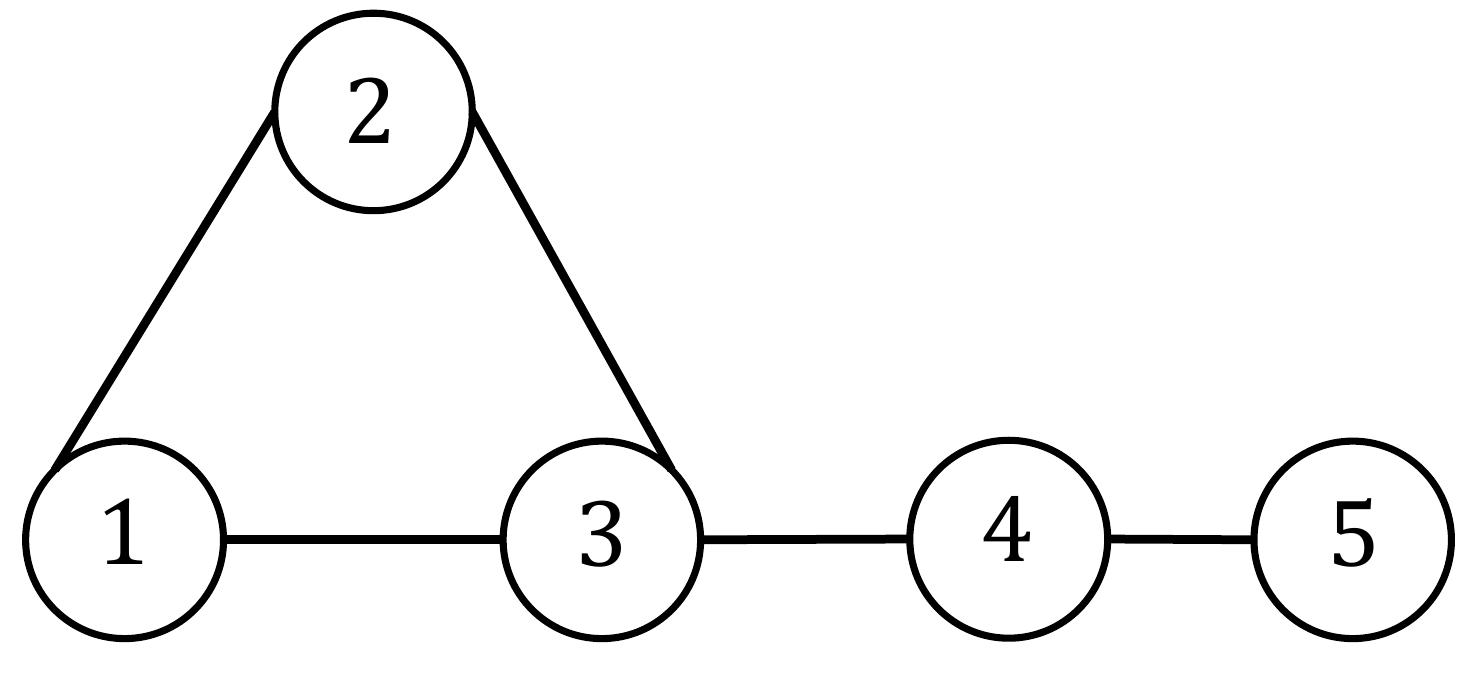} &
		\includegraphics[width=0.3\columnwidth]{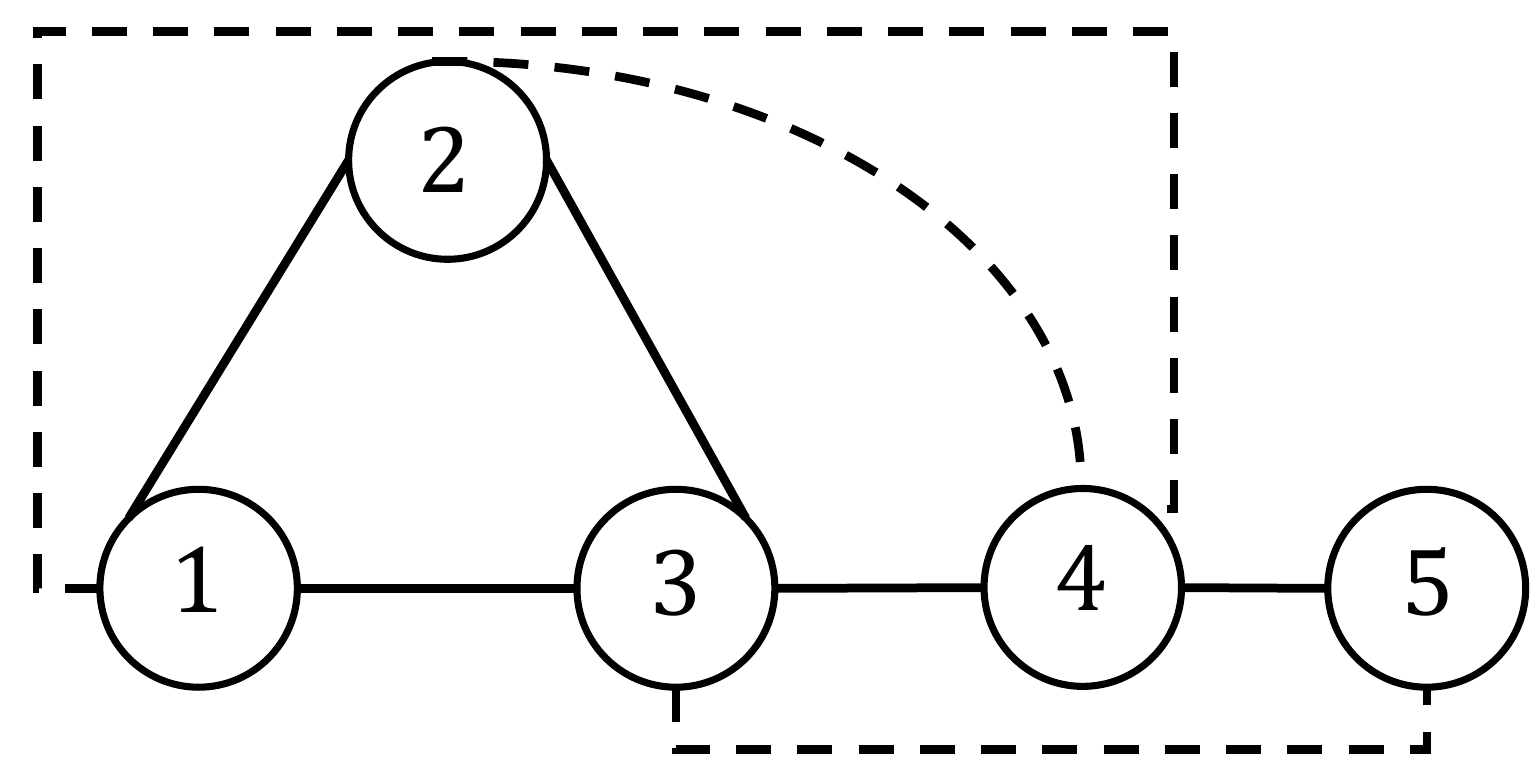}\\
		(a) & (b)&(c)
	\end{tabular}
	\caption{(a) Generative graph of $5$ node network undergoing consensus dynamics, (b) associated network topology, (c) reconstructed topology of the $5$ node network of Fig. \ref{fig:example_5node}(a) obtained using multivariate Wiener filtering with $10^7$ samples from each node. The dashed links are the spurious links due to spouse relationship.}
		\label{fig:example_5node} 
\end{figure}

\begin{figure}
\centering
  \begin{subfigure}[b]{0.5\textwidth}
  \centering
  \includegraphics[width=0.7\columnwidth]{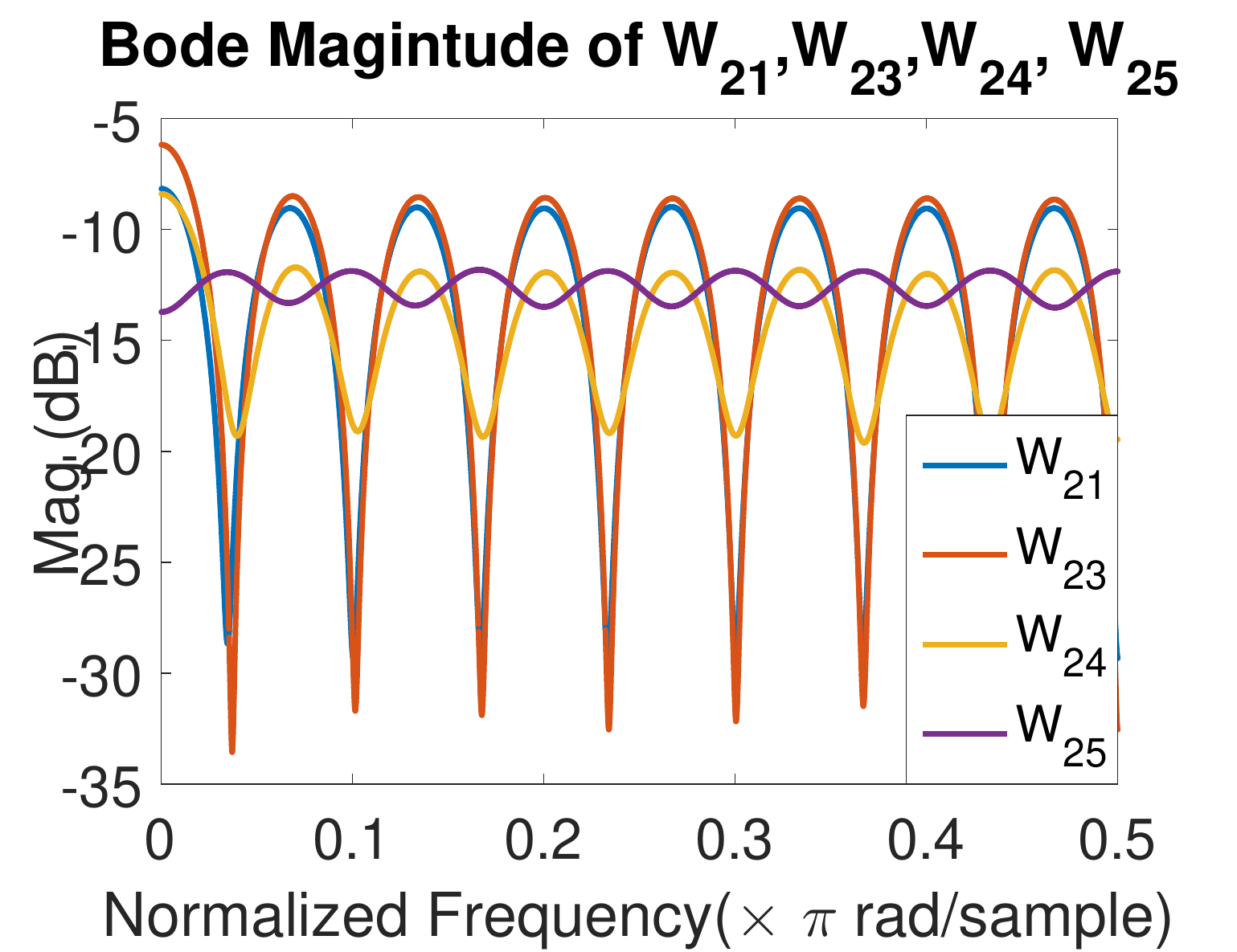}%
  \caption{}
  \end{subfigure}
\begin{subfigure}[b]{0.5\textwidth}
\centering
  \includegraphics[width=0.7\columnwidth]{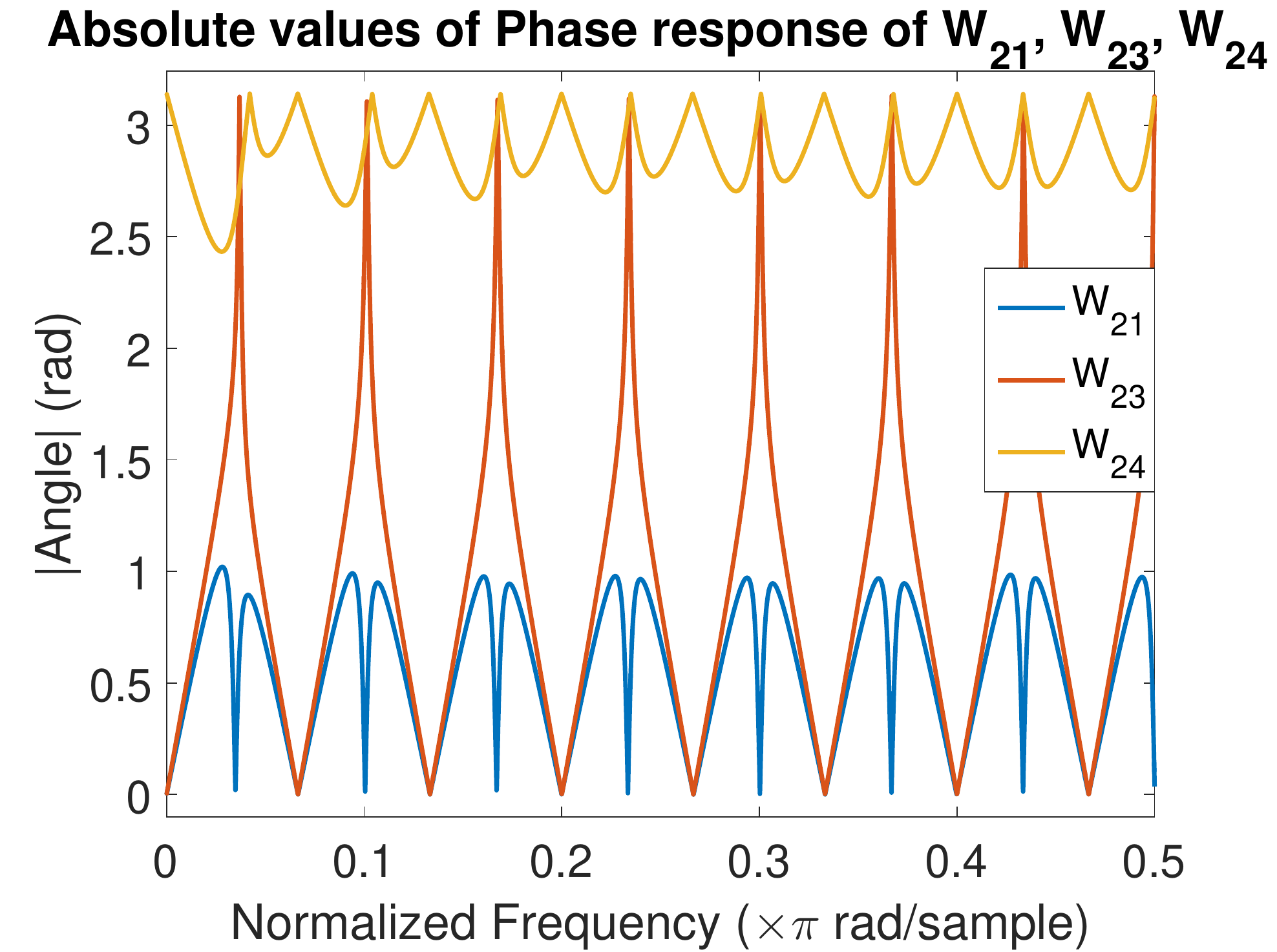}%
\caption{}
\end{subfigure}
\caption{(a) Bode magnitude plot of $W_{21}(z),W_{23}(z),W_{24}(z), W_{25}(z)$, (b) absolute values of phase response of $W_{21}(z)$,$W_{23}(z)$,$W_{24}(z)$, $W_{25}(z)$. $W_{24}(z)$ has a phase response in the vicinity of $\pi$ for all frequencies, hence, is eliminated by the pruning step.}
\label{fig:mag_5node} \label{fig:phase_5node}
\end{figure}

Here, the performance of Algorithm $2$ in estimating the topology of a network of agents undergoing consensus iterations is demonstrated on a $5$ node network depicted in Fig. \ref{fig:example_5node}(a). We present two set of results, first based on MATLAB simulations, and second based on experimental results on a network of Raspberry Pis. For our simulations, the receiver noise at each node is considered to be white. The trends from the nodal measurement time series are removed using the \lq detrend\rq\ function in MATLAB. The reconstructed topology using Algorithm $1$ with $10^7$ samples from each node is shown in Fig. \ref{fig:example_5node}(c), where, the dashed edges denote the spurious links recovered. Fig. \ref{fig:mag_5node}(a) and Fig. \ref{fig:phase_5node}(b) show the frequency response of Wiener filters between node $2$ and all other nodes in Fig. \ref{fig:example_5node}(a) that are derived using Algorithm $2$ with $10^7$ samples for each node. It is clear from Fig. \ref{fig:mag_5node}(a) that the magnitude of the filter $W_{25}$ is small across the frequency range and thus it can be concluded that there exist no edge between nodes $2$ and $5$. Using the pruning step, the absolute values of the phase response of $W_{21}(z), W_{23}(z), W_{24}(z)$ are analyzed as shown in Fig. \ref{fig:phase_5node}(b). Here, the phase of the filter $W_{24}$ remains close to $\pi$ throughout the frequency range. It follows from Theorem~\ref{thm:piresult} that the edge between $2$ and $4$ is spurious (phase response being close to $\pi$).

\begin{figure}
\centering
  \begin{subfigure}[b]{0.5\textwidth}
  \centering
  \includegraphics[width=0.6\columnwidth]{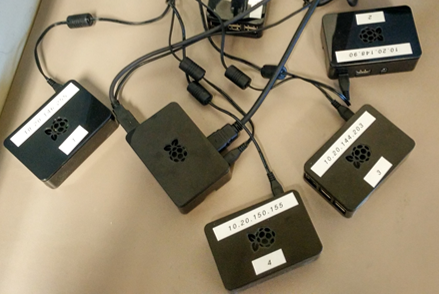}%
  \caption{}
  \end{subfigure}
\begin{subfigure}[b]{0.5\textwidth}
\centering
  \includegraphics[width=0.7\columnwidth]{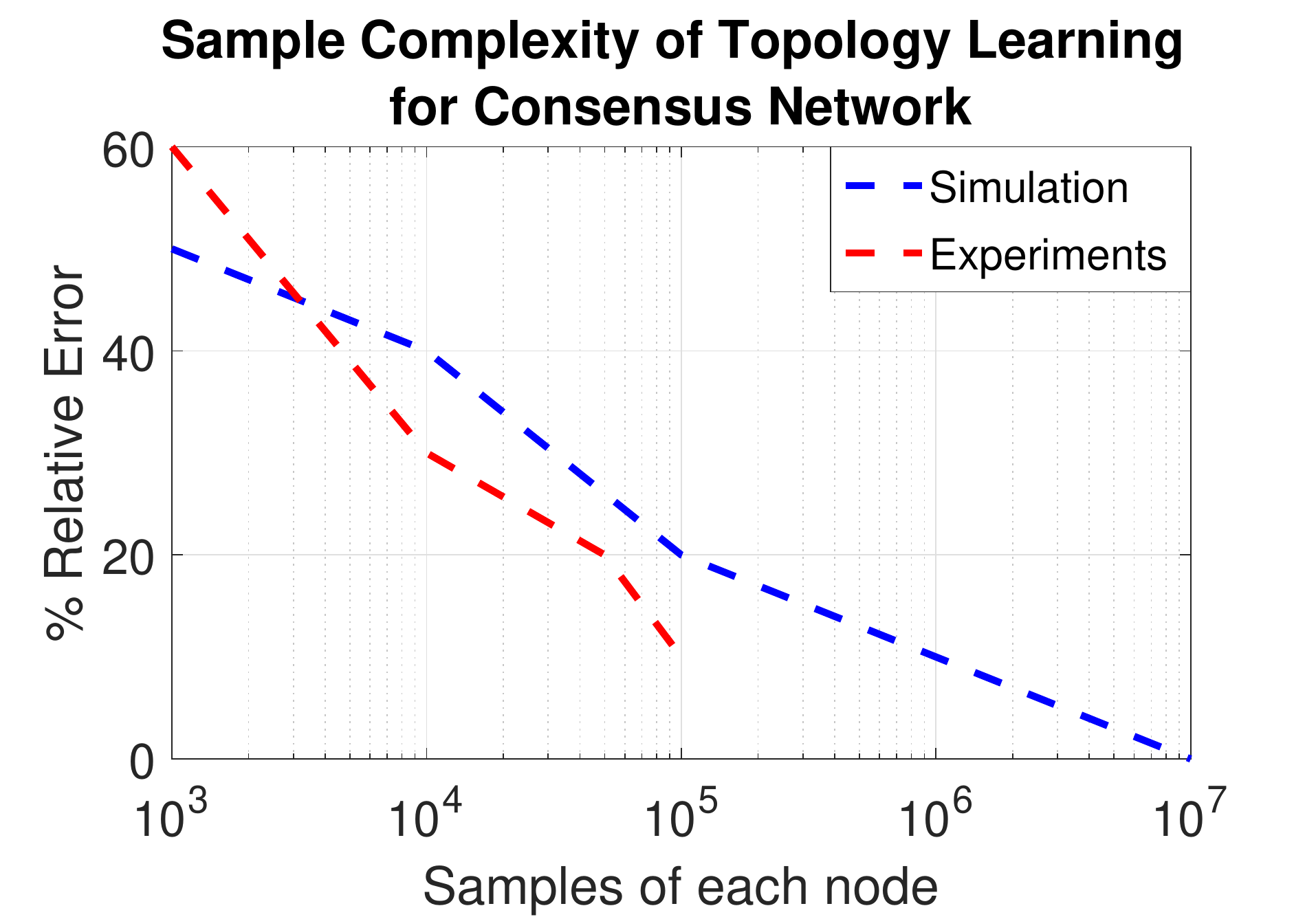}%
\caption{}
\end{subfigure}
\caption{(a) Experimental setup of $5$ Raspberry Pi units interacting through wifi according to consensus dynamics with the interaction topology being the undirected graph in Fig. \ref{fig:example_5node}, (b) error percentage variation with number of samples per node in simulation as well as experiments.}
	\label{fig:pi_units}
	\label{fig:error}
\end{figure}

Fig. \ref{fig:pi_units}(a) shows the experimental setup, which consists of five Raspberry Pi \cite{upton2014raspberry} units that interact according to consensus dynamics with the interaction topology described by Fig. \ref{fig:example_5node}(a). The details of the experimental platform can be found in \cite{prakash2016max}. The relative error (false negatives and false positives over the number of true edges) percentage of Algorithm $2$ with respect to number of samples for simulations as well as experiments is shown in Fig. \ref{fig:error}(b). It is seen that as the number of samples per node increases, the error decreases.

\vspace{-1 cm}
\vspace{1 cm}
\subsection{Validation on Power Distribution Network}
\begin{figure}
\centering
  \begin{subfigure}[b]{0.5\textwidth}
  \centering
  \includegraphics[width=0.7\columnwidth]{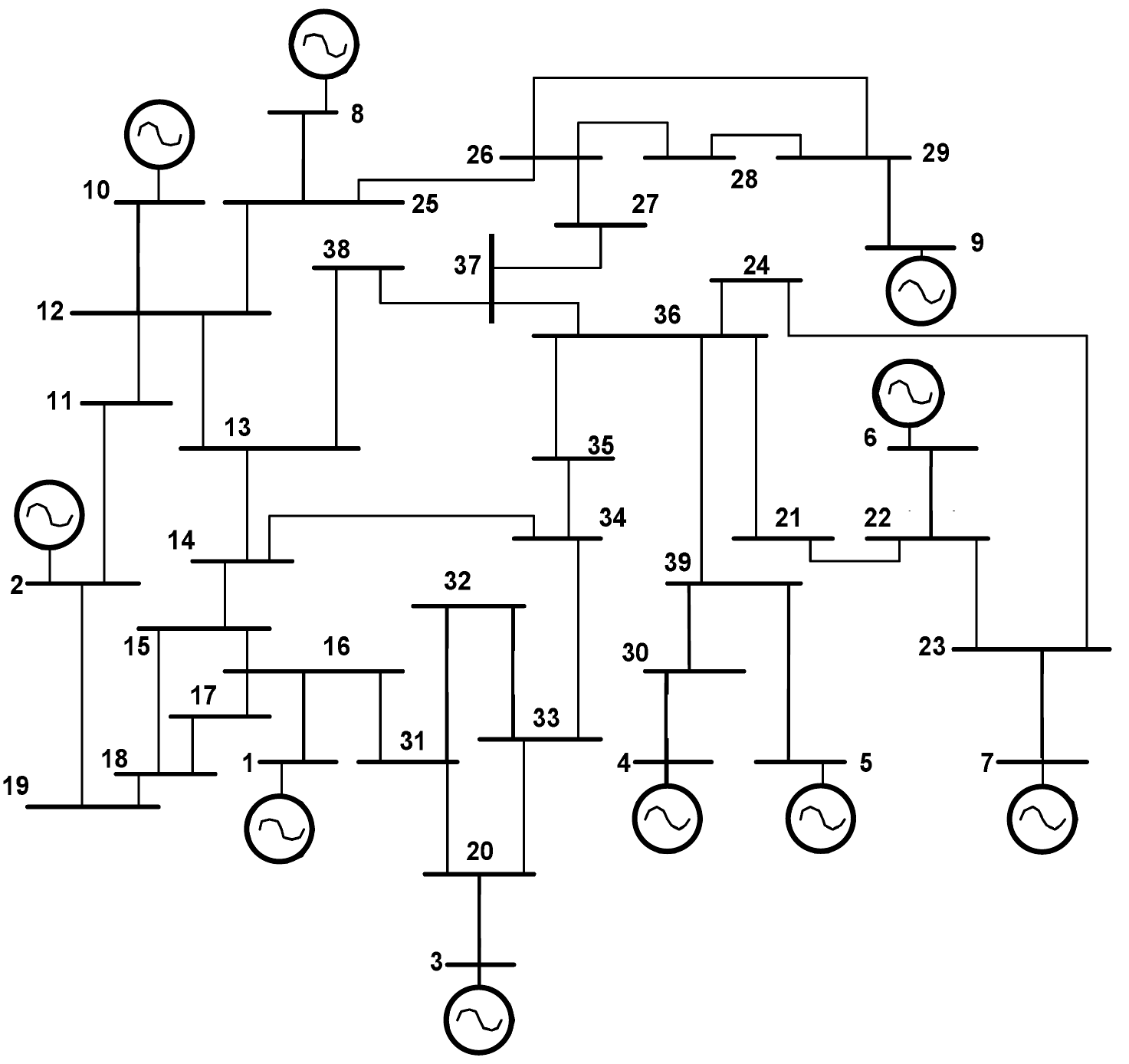}%
  \caption{}
  \end{subfigure}
\begin{subfigure}[b]{0.5\textwidth}
\centering
  \includegraphics[width=0.5\columnwidth]{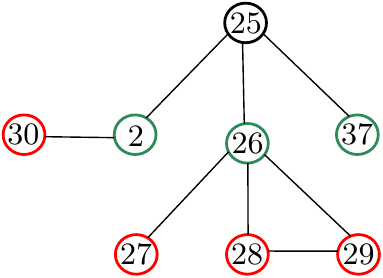}%
\caption{}
\end{subfigure}
\caption{(a) IEEE 39 bus system with generators at $10$ buses \cite{athay1979practical} (b) The neighbors (green nodes) and strict two-hop neighbors (red nodes) of node $25$ in the IEEE $39$ bus system.}
\label{fig:ieee39} \label{fig:ieee_39_node25}
\end{figure}

\begin{figure}
\centering
  \begin{subfigure}[b]{0.5\textwidth}
  \centering
  \includegraphics[width=0.8\columnwidth]{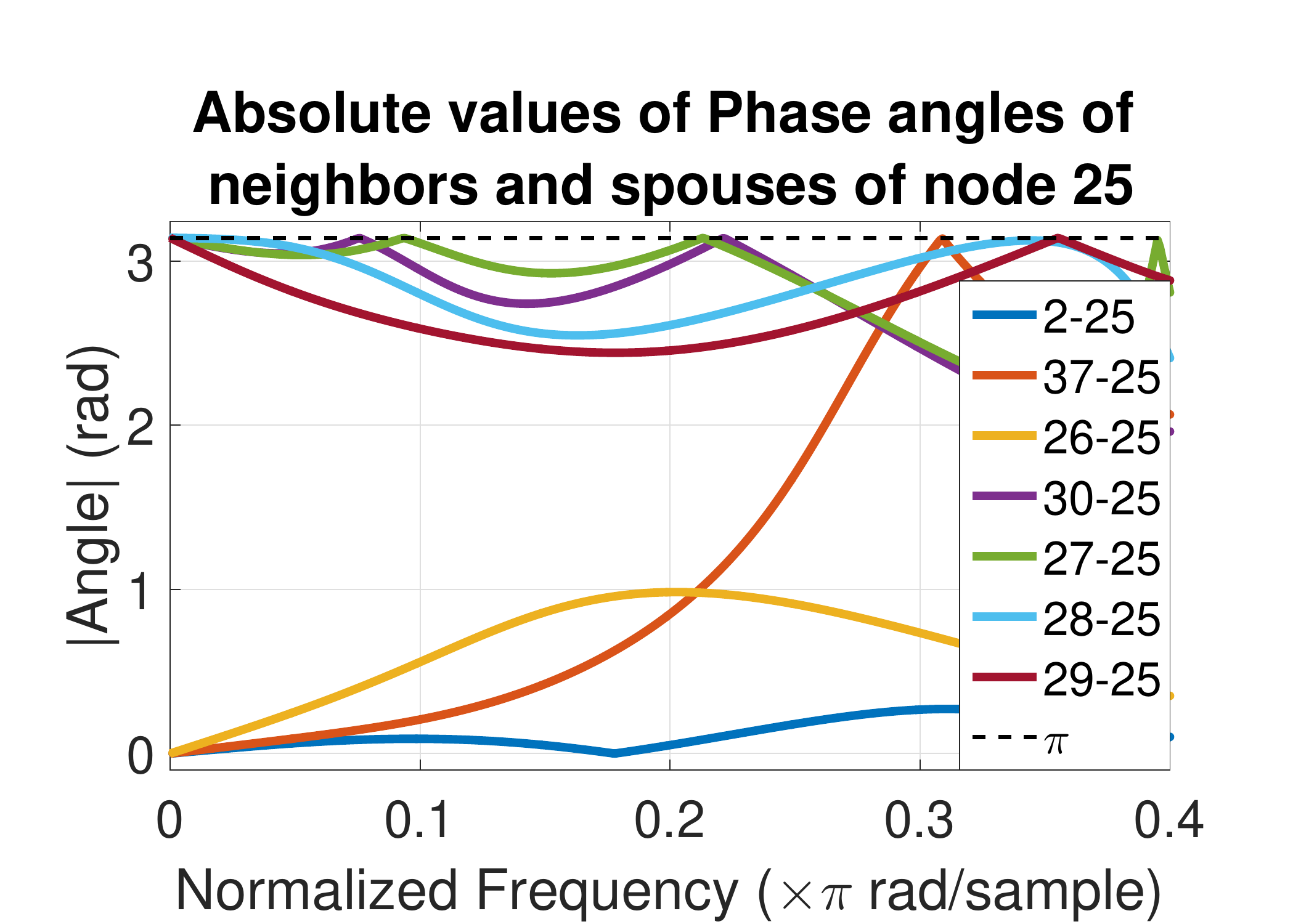}%
  \caption{}
  \end{subfigure}
\begin{subfigure}[b]{0.5\textwidth}
\centering
  \includegraphics[width=0.8\columnwidth]{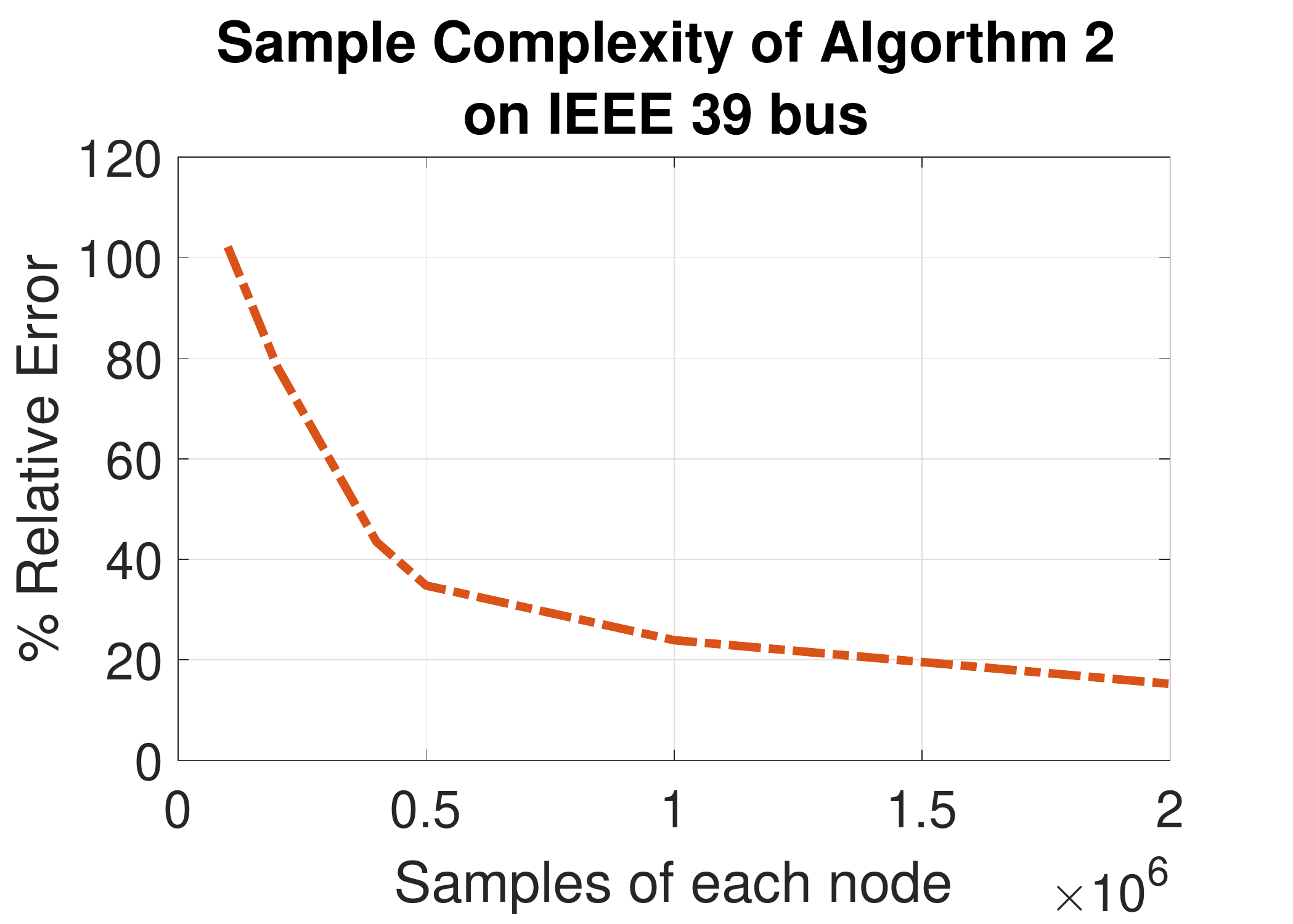}%
\caption{}
\end{subfigure}
\caption{(a) \small{Absolute values of the phase response of the Wiener filters between node $25$ and its two-hop neighborhood in the IEEE $39$ bus system. The phase response begins from $0$ rad for all three neighbors and from $\pi$ for all spouses of node $25$. } (b) \small{Relative error percentage of Algorithm 2 with samples per node for IEEE 39 bus systems.}}
\label{fig:phase_1hops} \label{fig:error_proportion}
\end{figure}

\vspace{-0.5 cm}
In this section, we demonstrate the effectiveness of Algorithm $2$ on the IEEE 39 bus power distribution network \cite{athay1979practical} shown in Fig.~\ref{fig:ieee39}(a) with network dynamics as described by (\ref{swingmatrix}). Here $p_i$ are modeled as filtered white Gaussian noise (colored noise unlike iid Gaussian in \cite{dekapscc}) to generate time series data for evaluation of the proposed algorithm. The output at each node is sampled at $0.01 s$.
Consider the neighbors (green) and two-hop neighbors (red) of node $25$ in the IEEE 39 bus system as shown in Fig~\ref{fig:ieee_39_node25}. Application of steps $1$ to $9$ of Algorithm $1$ with a threshold $\rho$ of $10^{-5}$ in step \ref{step_nonzero} and $6.5 \times 10^6$ phase samples per node leads to edges between node $25$ and all nodes in Fig.~\ref{fig:ieee_39_node25}(b). The absolute values of the phase response of the multivariate Wiener filters for node $25$ and the nodes in its two-hop neighborhood are shown in Fig.~\ref{fig:phase_1hops}(a). It is seen that the phase response of the Wiener filters corresponding to the nodes two hops away are close to $\pi$ rad, while that of the neighbor nodes start from $0$ rad. Thus using the pruning step, all the two-hop neighbors can be removed, recovering the true physical topology. The relative error percentage in topology estimation for the IEEE 39 bus system as a function of sample size is shown in Fig.~\ref{fig:error_proportion}(b). The threshold $\rho$ was chosen as $10^{-3}$ while $\tau = 0.2 \pi$. In many cases, pruning step eliminated $58 \%$ of the total false positive edges obtained after step $9$ of Algorithm $2$.

\subsection{Comparative study and validation on Thermal Dynamics of Building}
\vspace{-0.5 cm}

\begin{figure}[tb]
	\centering
	\begin{tabular}{cc}
		\includegraphics[width=0.5\columnwidth, height = 3 cm]{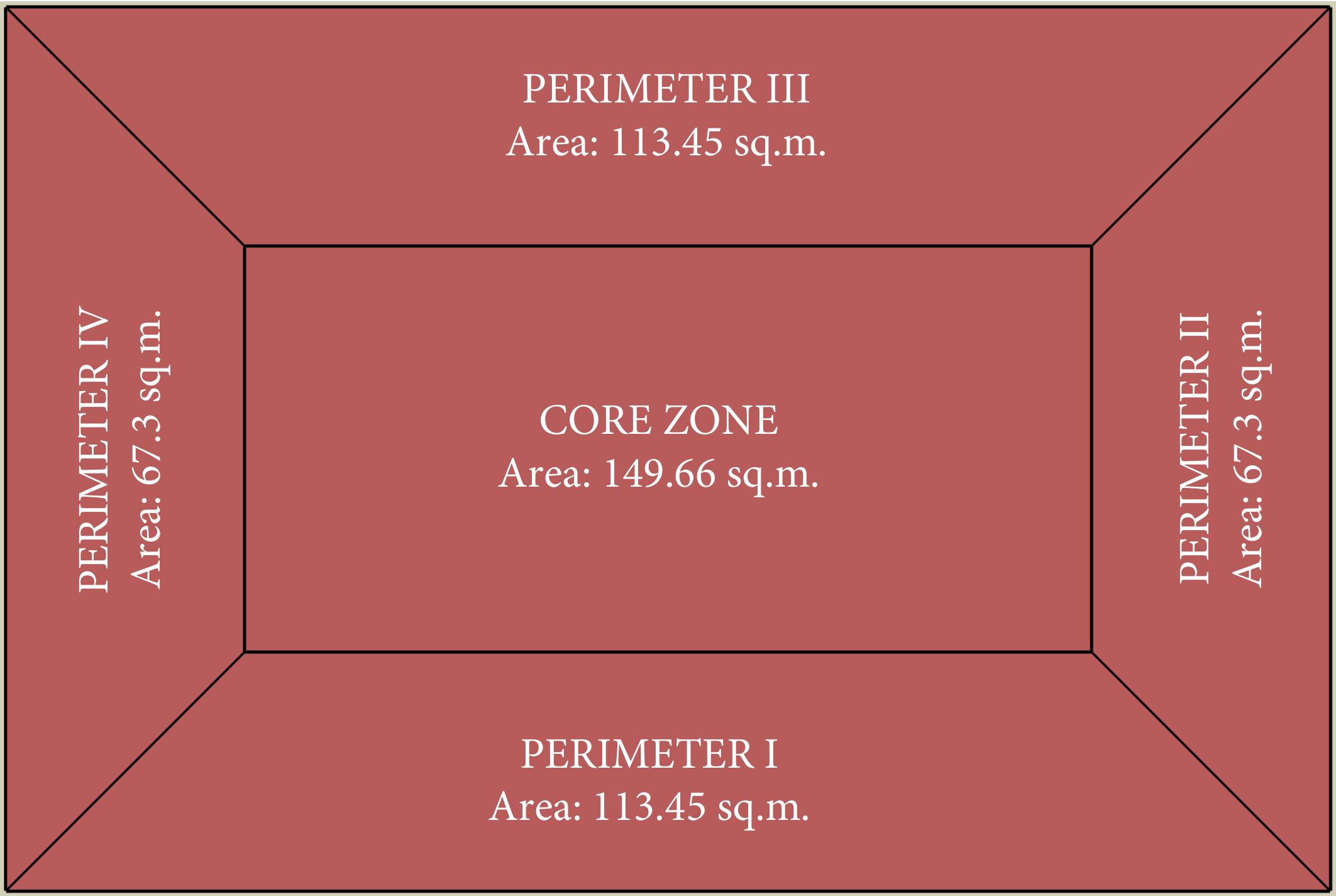}&
		\includegraphics[width=0.5\columnwidth, height = 3 cm]{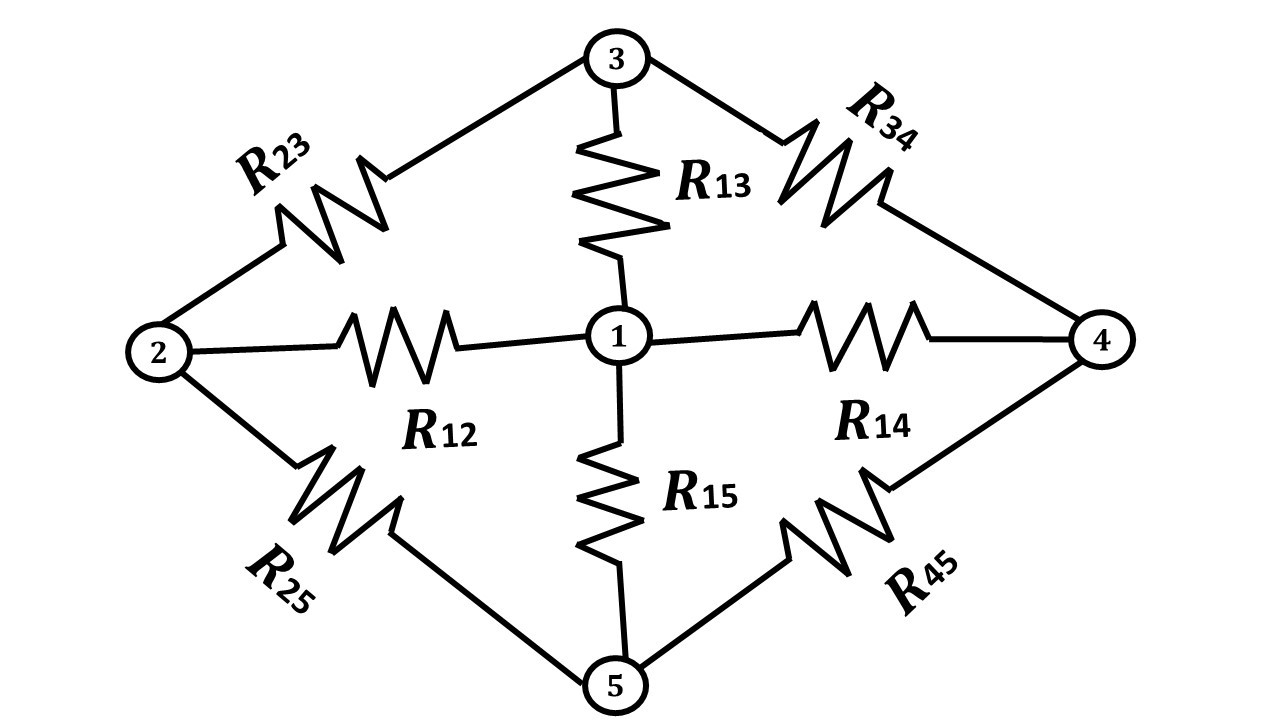} \\
		(a) & (b)\\
	\end{tabular}
	\caption{(a) Topview of EnergyPlus building model consisting of $5$ zones, where, the height of the building is 3.05 m, (b) Thermal resistor network of the building with thermal capacitance at each node. Node 1 corresponds to core zone and rest of the nodes are referred to as perimeter zones}\label{fig:energyplus_5node}
\end{figure}
In this section, we will estimate the interaction topology amongst zonal temperatures in a building using Algorithm 2 and also highlight the limitations in inference of three node cycles using static approaches. Here, we illustrate that by applying Algorithm $2$ to the temperature data of a five zone office building, the true topology is recovered exactly. The building envelope is created in Google SketchUp Make 2017 \cite{ellis2008energy} (see Fig. \ref{fig:energyplus_5node}) and its RC network model in Fig. \ref{fig:energyplus_5node}(b). The temperature data is generated using EnergyPlus \cite{crawley2000energyplus}. EnergyPlus solves the nonlinear energy balance equation where the heat transfer coefficients are functions of temperature (See equation \ref{heatequation} in Appendix $A$). For more details on EnergyPlus see Section $A$ in Appendix. The building is located in Minneapolis, Minnesota, USA and the weather file used in EnergyPlus is obtained from \url{https://energyplus.net/weather}. The exogenous inputs to the building are heat gains from lights, electrical equipments and people. For our simulations, we consider the electrical and lighting loads as time-correlated wide sense stationary processes. The correlated inputs are generated by filtering white Gaussian noise through 1D digital filter in MATLAB.

\begin{figure}[tb]
	\centering
	\begin{tabular}{cc}
		\includegraphics[width=0.95\columnwidth, height = 5.2 cm]{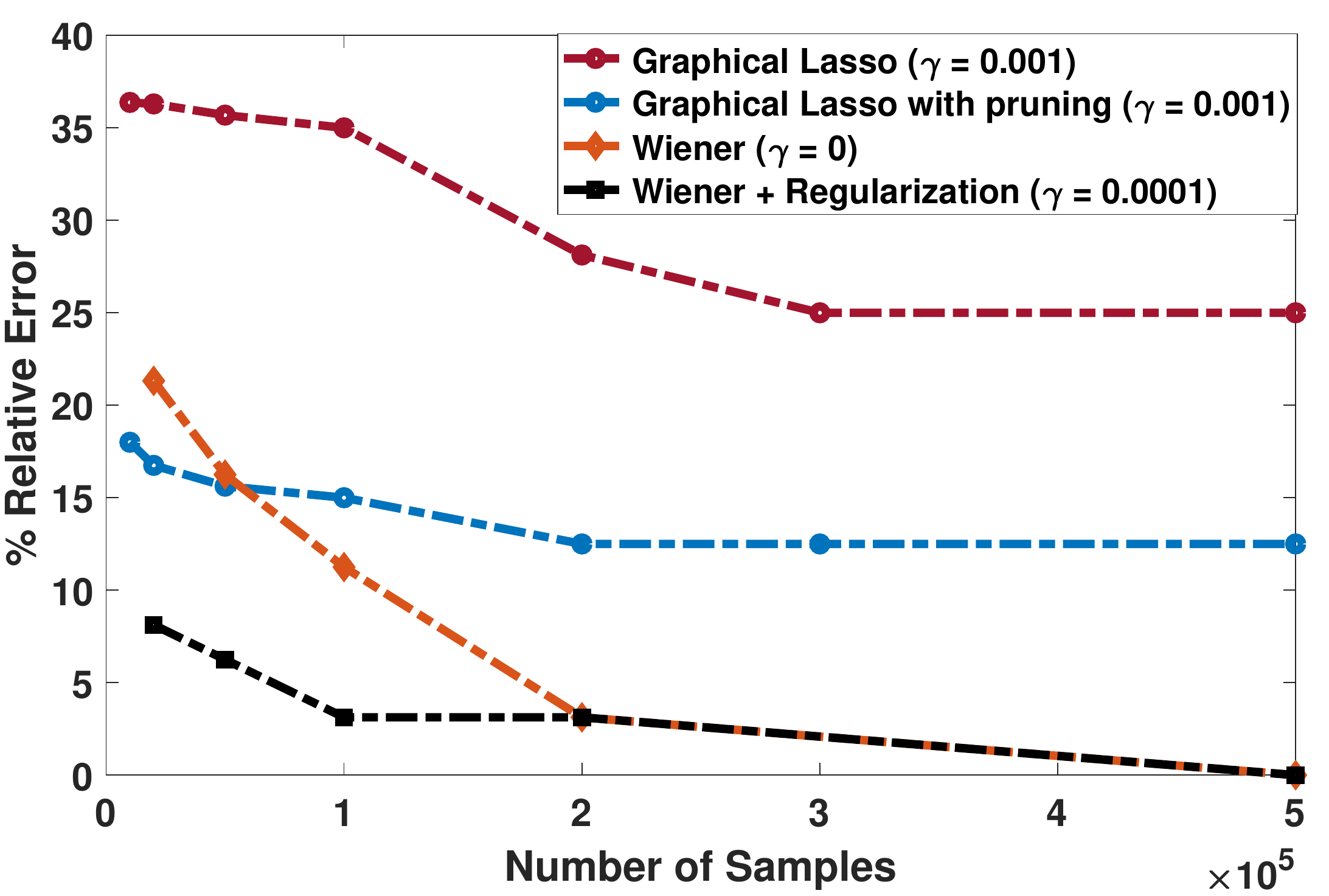}\\
	\end{tabular}
\caption{Error percentage variation with number of samples per node for Algorithm $2$ (with and without regularizers) when inputs are WSS \label{fig:errorplot}}
\end{figure}

This temperature data is obtained with one minute granularity and used for topology inference. Here, we perform a comparative study of four algorithms: graphical lasso \cite{friedman2008sparse} (random variable framework for Gaussian graphical models, see Appendix $B$ for details), graphical lasso with sign based pruning \cite{saverio}, \cite{dekairep}(see Appendix $C$ for details), Algorithm 2 with Wiener filtering computation according to (\ref{wiener}) and Algorithm 2 with Wiener filtering computation with regularizer (group lasso \cite{friedman2001elements}) as described in (\ref{regularized}) below. The motivation behind introducing regularization in the Wiener filter optimization problem is to improve the performance of Algorithm $2$ in low sample regime. It is seen in Fig. \ref{fig:errorplot} that the error in the low sample regime is about $20\%$ for Algorithm $2$ with Wiener filters computed using the approach described in (\ref{wiener2}). Regularizers are commonly used to improve the performance of inference algorithms primarily in the high dimensional setting, where, number of nodes are large and samples are limited \cite{meinshausen2006high,friedman2008sparse}. Here, we introduce the group lasso regularizer \cite{friedman2001elements} to enforce sparsity and reduce over-fitting in the case of availability of limited number of samples per node.

\textbf{Wiener filter with regularization: }\label{WF}
Here, we present a method  that provides estimates of the optimal Wiener filter which is well suited for scenarios when data-records are short. In order to account for limited samples of measurements at each node, a regularized version of multivariate Wiener filtering in (\ref{wiener}) is obtained by minimizing the following objective function:
\begin{align}\label{regularized}
\{h_{ji,\gamma}\}&={\arg \inf_{{\{h_{ji}\}}_{i=1,...,m,i\neq j}}}~~ \nonumber\\
&\mathbb{E}(x_j(k)-
		\smashoperator[lr]{\sum_{i=1,i\neq j}^{m}}~\sum_{L=-F}^{F}h_{ji}^{L}x_i(k-L))^2 +\gamma\smashoperator[lr]{\sum_{i=1,i\neq j}^{m}}
	\|h_{ji}\|_2.
\end{align}
Here $\gamma \geq 0$ is the regularization parameter and $h_{ji}:=[h_{ji}^{-F},...,h_{ji}^{0},...,h_{ji}^{F}]$. The regularized Wiener filter is given as,
\begin{align}\label{regularfilter}
	  W^\gamma_{ji}(z) = \sum_{L=-F}^{F}h_{ji,\gamma}^{L}z^{-L}.
	\end{align}
Using regularized Wiener filters in Algorithm $2$, the relative errors for the topology inference of the thermal dynamics of the building described above is shown in Figure \ref{fig:errorplot}. It is seen that in the low sample regime, the error in inference reduces by about half due to use of regularization in Wiener filter computations. The relative error percentage with number of samples for the graphical lasso based algorithms is shown in Fig.\ref{fig:errorplot}. It is seen that, the graphical lasso and graphical lasso with pruning step is unable to recover the exact topology of the underlying RC network, even in the large sample limit. This is a limitation of the random variable framework for networks with three node undirected cycles, which was also highlighted in the introduction of this article. However, the time series based approach of Algorithm 2 with and without regularizer have zero error in topology inference in the large sample limit. This is attributed to the ability to analyze the Wiener filters at multiple frequencies in the time series framework as compared to only $\omega = 0$ in the random variable setting, which renders three node undirected cycles unidentifiable.

\section{Conclusion}
In this article, we presented an algorithm for exact topology inference of linear dynamical systems with a particular focus on physical flow networks. The flow conservation constraint provides a basis for pruning out spurious spouse links, which crop up in topology learning from time series measurements. We utilized the phase response of multivariate Wiener filters to distinguish between true and spurious edges. The algorithm proposed in this article can handle colored noise exogenous inputs, unlike the white noise assumption in some of the recent work on topology inference from time series measurements. Moreover, the proposed algorithm guarantees exact topology inference even in the presence of loops in the network. It is worth noting that, the proposed algorithm does not require any knowledge on statistics of the exogenous inputs or system parameters for topology inference. We believe that using phase response of filters to eliminate spurious links in topology learning can also be extended to other network topology learning approaches like directed information and spectral analysis, as connections of these approaches with the Wiener filtering approach is well understood \cite{etesami2014directed}, \cite{materassi2012problem}. The article, also highlighted the short comings of state of the art random variables approaches in dealing with networked dynamical systems, particularly in inferring three node cycles and the superiority of the algorithm proposed in such scenarios. We also illustrated the benefit of using regularizers in Wiener filter computations, for obtaining better topology inference performance in the low sample regime. This is particularly important for deploying the proposed algorithm for topology inference from finite size data windows of measurements across the network.

\begin{ack}
The authors S. Talukdar, H. Doddi, D. Materassi and M. V. Salapaka
acknowledge the support of ARPA-E for supporting this
research through the project titled \lq A Robust Distributed
Framework for Flexible Power Grids\rq \ via grant no. DEAR000071. Authors D. Deka and M. Chertkov acknowledge the support from the Department of Energy through the Grid Modernization Lab Consortium, and the Center for Non Linear Studies (CNLS) at Los Alamos.
\end{ack}

\bibliography{autosam}      



\appendix
\section{Overview of EnergyPlus}
EnergyPlus is a building energy simulation engine developed by U.S. Department of Energy. It is an open-source software that can be downloaded at \url{https://energyplus.net/downloads}. The inputs to the EnergyPlus is a text file with detailed information of the building structure, construction, equipment, location, orientation, weather details along with the occupancy, electrical, lighting schedules. It is a sophisticated simulation tol for thermal analysis of building.

EnergyPlus assumes each thermal zone as a single node and solves the heat balance equations to arrive at the thermal zone temperatures developed in the building and its power consumption. The heat balance equation for a zone \cite{doe2010energyplus} is given by:
\begin{align}\label{heatequation}
&\mathcal{C}_{z}\frac{dT_z}{dt} = \sum\limits_{i=1}^{N_{sl}}\Dot{Q_{i}} +\sum\limits_{i=1}^{N_{surf}}h_{i}A_{i}(T_{si}-T_{z}) \\
&~+\sum\limits_{i=1}^{N_{zones}}\Dot{m_{i}}\mathcal{C}_{p}(T_{zi}-T_{z})+\Dot{m}_{inf}\mathcal{C}_{p}(T_{\infty}-T_{z})+\Dot{Q}_{sys}\nonumber\\
&\text{where~} \mathcal{C}_z\frac{dT_z}{dt} \text{~is the energy stored in zone air,}\nonumber\\
&\sum\limits_{i=1}^{N_{sl}}\Dot{Q_{i}} \text{~is internal convective load,}\nonumber\\
&\sum\limits_{i=1}^{N_{surf}}h_{i}A_{i}(T_{si}-T_{z}) \text{~is surface convective heat transfer,} \nonumber\\
&\sum\limits_{i=1}^{N_{zones}}\Dot{m_{i}}\mathcal{C}_{p}(T_{zi}-T_{z}) \text{~is interzone heat transfer, and}\nonumber\\
&\Dot{m}_{{inf}}\mathcal{C}_{p}(T_{\infty}-T_{z}) \text{~is heat transfer by air infiltration, and, }\nonumber\\
&\Dot{Q}_{{sys}} \text{~is air systems output.}\nonumber\\
& \text{Note that $h_i$ is a non-linear function of temperatures.}\nonumber
\end{align}

\section{Graphical Lasso}
In the random variable framework \cite{friedman2008sparse}, consider $N$ independent and identically distributed random variables of a dimension $p$, with mean $\mu$ and covariance $\Sigma$. Let $\Theta = \Sigma^{-1}$, $\rho$ is the regularization parameter and $S$ be the empirical covariance matrix. Then the maximum likelihood estimator of $\Theta$ based on sparsity constraints is given by,
\begin{align}
  \Theta^{*} = \argmax_\Theta {log det \Theta -tr(S\Theta) -\rho \|\Theta\|_1 }
  \label{graphLasso_eq}
\end{align}
 Please see the Algorithm \ref{Glasso} for step by step procedure for topology reconstruction.
\begin{algorithm}
\caption{Topology Learning using Graphical Lasso}
\label{Glasso}
\textbf{Input:} nodal time samples $x_i(k)$ for nodes $i \in \{1,2,...,n\}$ in the generative graph $\mathcal{G}$, thresholds $\epsilon$\\
\textbf{Output:} Estimate of Edges $\bar{{\mathcal{E}}}_T$ in the topology of $\mathcal{G}$\\
\begin{algorithmic}[1]
\ForAll{$j \in \{1,2,...,n\}$}\label{step1_a}
\State Compute inverse covariance matrix $\Theta^{*}$ from \ref{graphLasso_eq} \label{step1_a1}
\EndFor
\State Edge set ${\mathcal{E}}^{w} \gets \{\}$
\ForAll{$i,j \in \{1,2,...,n\}, i\neq j$}
\If{${|\Theta^*}_{ij}| > \epsilon$}\label{step_nonzero}
\State $\mathcal{E}^w \gets \mathcal{E}^w \cup \{(i,j)\}$
\EndIf
\EndFor\label{step1_b}
\end{algorithmic}
\end{algorithm}
\vspace{4 cm}
\section{Graphical lasso with sign based pruning}
After estimating the inverse covariance matrix $\Theta^{*}$ as described above, the steps for identifying the neighbors and pruning out the spurious links are described here based on the. description in \cite{saverio}, \cite{dekairep}.
\begin{algorithm}
\caption{Topology Learning using Graphical Lasso with sign based pruning}
\label{Glasso_sign}
\textbf{Input:} nodal time samples $x_i(k)$ for nodes $i \in \{1,2,...,n\}$ in the generative graph $\mathcal{G}$, thresholds $\epsilon$\\
\textbf{Output:} Estimate of Edges $\bar{{\mathcal{E}}}_T$ in the topology of $\mathcal{G}$\\
\begin{algorithmic}[1]
\ForAll{$j \in \{1,2,...,n\}$}\label{step1_a}
\State Compute $\Theta^{*}$ from \ref{graphLasso_eq} \label{step1_a1}
\EndFor
\State Edge set ${\mathcal{E}}^{w} \gets \{\}$
\ForAll{$i,j \in \{1,2,...,n\}, i\neq j$}
\If{${|\Theta^*}_{ij}| > \epsilon$}\label{step_nonzero}
\State $\mathcal{E}^w \gets \mathcal{E}^w \cup \{(i,j)\}$
\EndIf
\EndFor\label{step1_b}
\State Edge set $\bar{\mathcal{E}}_T \gets {\mathcal{E}}^{w}$ \label{step2_a}
\ForAll{$i,j \in \{1,2,...,n\}, i\neq j$}
\If{$\Theta^*_{ij} > \epsilon,$}
\State $\bar{\mathcal{E}}_T \gets \bar{\mathcal{E}} - \{(i,j)\}$
\EndIf
\EndFor \label{step2_b}
\end{algorithmic}
\end{algorithm}

\end{document}